\documentclass{amsart}
\date{\today}
\def\M{\|}
\newcommand{\R}{\mathbb{R}}
\newcommand{\E}{\mathbb{E}}
\newcommand{\N}{\mathbb{N}}
\newcommand{\C}{\mathbb{C}}
\newcommand{\Z}{\mathbb{Z}}
\newcommand{\p}{\mathbb{P}}
\newcommand{\Cal}{\mathcal}

\newtheorem{Theorem}{Theorem}[section]
\newtheorem{Proposition}[Theorem]{Proposition}
\newtheorem{Corollary}[Theorem]{Corollary}
\newtheorem{Lemma}[Theorem]{Lemma}

\newtheorem{Example}[Theorem]{Example}
\newcommand{\dist}{\mbox{\rm dist}}

\renewcommand{\epsilon}{\varepsilon}

\begin{document}
\setcounter{section}{0}
\renewcommand{\theequation}{\arabic{section}.\arabic{equation}}
\newcounter{letters}

\title[Bernoulli-Anderson Localization]{Localization for One Dimensional,
Continuum, Bernoulli-Anderson Models}

\author[D.\ Damanik]{David Damanik}
\address{Department of Mathematics, University of California at Irvine, CA 92697-3875, USA, partially supported by the German Academic Exchange Service through HSP~III (Postdoktoranden)}

\author[R.\ Sims]{Robert Sims}
\address{Department of Mathematics, University of
Alabama at Birmingham, AL 35394-1170, USA, partially supported by
NSF grant DMS-9706076}

\author[G.\ Stolz]{G\"unter Stolz}
\address{Department of
Mathematics, University of Alabama at Birmingham, AL 35394-1170,
USA, partially supported by NSF grants DMS-9706076 and
DMS-0070343}

\maketitle

\begin{abstract}

We use scattering theoretic methods to prove strong dynamical and
exponential localization for one dimensional, continuum,
Anderson-type models with singular distributions; in particular
the case of a Bernoulli distribution is covered. The operators we
consider model alloys composed of at least two distinct types of
randomly dispersed atoms. Our main tools are the reflection and
transmission coefficients for compactly supported single site
perturbations of a periodic background which we use to verify the
necessary hypotheses of multi-scale analysis. We show that
non-reflectionless single sites lead to a discrete set of
exceptional energies away from which localization occurs.

\end{abstract}


\section{Introduction}
Perhaps the most studied and best understood type of random
operators which are used to describe spectral and transport
properties of disordered media are the Anderson models, whether
one considers their original discrete version in $\ell^2( \Z^d)$
or their continuum analogs. They describe materials of alloy type,
i.e. a structure in which single site potentials are centered at
the points of a regular lattice and then multiplied by random
coupling constants, modelling the differing nuclear charges of the
alloy's component materials. It is therefore physically most
relevant to study the case where the coupling constants take on
only finitely many values. The special case of a two-component
alloy, i.e. coupling constants given by Bernoulli random
variables, has been dubbed the Bernoulli-Anderson model.
Unfortunately, most of the rigorous results on Anderson models
require more regularity of the probability distributions governing
the coupling constants. Results on exponential localization and,
more recently, dynamical localization have usually been proven
under the assumption that the distribution is absolutely
continuous or, at least, has an absolutely continuous component,
see \cite{Carmona/Lacroix} and \cite{Figotin/Pastur} for results
from the 80s, and \cite{Aizenman/Molchanov, Bi/Ger1,
Combes/Hislop, FK, Kirsch/Stollmann/Stolz, Klopp, Stolz:Anderson}
for some of the more recent contributions. In dimension $d>1$ the
best results with respect to weak regularity are due to Carmona,
Klein, and Martinelli \cite{CKM} for the discrete case and to
Stollmann \cite{PSW} and Damanik and Stollmann \cite{DS} for
continuum models, where localization properties are shown under
merely assuming h\"{o}lder-continuity of the distribution.

Discrete distributions and, in particular, the Bernoulli-Anderson
model, are currently only accessible in $d=1$. For the discrete,
one-dimensional Anderson model
\begin{equation} \label{discreteam}
(h_{\omega}u)(n) = u(n+1)+u(n-1)+q_n( \omega)u(n), \ \ n \in \Z,
\end{equation}
exponential localization has been proven at all energies for
arbitrary, non-trivial distribution (i.e. support containing more
than one point) of the independent, identically distributed
(i.i.d.) random variables $q_n$. This was first proven by Carmona,
Klein, and Martinelli \cite{CKM} and later by Shubin, Vakilian,
and Wolff \cite{SVW} with a different approach which, in turn, is
based on results from \cite{Camp+}. An interesting recent paper by
De Bi\`{e}vre and Germinet \cite{Bi/Ger2} studies the so-called
{\it random dimer model}. This is the case where in $(
\ref{discreteam})$ one chooses $q_{2n+1}( \omega)=q_{2n}( \omega)$
and the $q_{2n}$, $n \in \Z$ as i.i.d. random variables with only
two values $\pm q$ for some $q>0$. \cite{Bi/Ger2} establishes
dynamical localization for the random dimer model in compact
intervals away from certain critical energies, where the Lyapunov
exponent vanishes and delocalized states exist.

In this paper we will prove localization, exponential and
dynamical, for {\it continuous} one-dimensional Anderson models
with arbitrary non-trivial distribution. Let
\begin{equation} \label{contam}
H_{\omega} = - \frac{d^2}{dx^2} + V_{{\rm per}} + V_{\omega}, \ \ \omega
\in \Omega,
\end{equation}
with
\begin{equation} \label{apot}
V_{\omega}(x) = \sum_{n \in \Z} q_n( \omega)f(x-n).
\end{equation}
We assume that the background potential $V_{{\rm per}}$ has period $1$,
is real-valued and locally in $L^1$. The single site potential $f
\in L^1$ is real-valued, supported in $[-1/2,1/2]$, and not $0$
(in $L^1$-sense). The coupling constants $q_n$ are i.i.d. random
variables on a complete probability space $\Omega$. We assume that
the support of their common distribution $\mu$ is bounded and
non-trivial, i.e.\ contains at least two points. Note that without
restriction we may assume that supp$(\mu)$ contains $0$ and $1$
(if $\{a,b\} \subset \mbox{supp}(\mu)$, then replace $V_{{\rm per}}$ by
$V_{{\rm per}} + a\sum_n f(\cdot-n)$ and $f$ by $(b-a)f$). We will do
this in all our proofs.

Under these assumptions the operators $H_{\omega}$ can be defined
either by form methods or via Sturm-Liouville theory and are
self-adjoint for every $\omega$. Representing, as usual, $\Omega$
as an infinite product and $q_n(\omega)$ as $\omega_n$, one easily
sees that
\begin{equation*}
H_{\theta_k\omega}f= \tau_k^*H_{\omega}\tau_kf \ \ \mbox{for } f
\in D(H_{\theta_k\omega}),
\end{equation*}
with $\tau_kf=f(\cdot -k)$ and the ergodic shifts $(\theta_k
\omega)_n= \omega_{n+k}$. Thus, e.g. \cite{Carmona/Lacroix},
$H_{\omega}$ has non-random almost sure spectrum $\Sigma$ and
spectral types $\Sigma_{ac}$, $\Sigma_{sc}$, and $\Sigma_{pp}$.
Our first main result is

\begin{Theorem}[Exponential Localization]\label{exploc}
Almost surely, the operator $H_{\omega}$ has pure point spectrum,
i.e. $\Sigma_{ac} = \Sigma_{sc} = \emptyset$, and all
eigenfunctions decay exponentially at $\pm \infty$.
\end{Theorem}

Our method to prove Theorem $\ref{exploc}$ is basically to adapt
the approach of \cite{CKM} to continuum models. We will, however,
use a variable-energy multiscale analysis, which was introduced
in \cite{VDK} for discrete models and later adapted to the
continuum, e.g. \cite{Stollmann, Bi/Ger1, FK}.

Recently, it has been demonstrated by Damanik and Stollmann
\cite{DS} that the variable-energy multiscale analysis implies
{\it strong dynamical localization}. Thus our method of proof will
also yield

\begin{Theorem}[Strong Dynamical Localization]\label{dynloc}
Let $H_{\omega}$ be defined as above. There exists a discrete set
$M \subset \R$ such that for every compact interval $I \subset \R
\setminus M$, and every compact set $K \subset \R$, and every
$p>0$,

\begin{equation} \label{sdl}
\E \left\{ \sup_{t>0} \M |X|^pe^{-itH_{\omega}}P_I(H_{\omega})
\chi_K \M \right\} < \infty,
\end{equation}
where $P_I$ is the spectral projection onto $I$.
\end{Theorem}

We note that the ``strong" in Theorem $\ref{dynloc}$ refers to the
ability to show finite expectation in $( \ref{sdl})$. This is
stronger than showing that the supremum in $( \ref{sdl})$ is
almost surely finite, which has also been used to describe
dynamical localization. For more details on dynamical localization
we refer to \cite{4author, Bi/Ger1, Aizen+, BFM, DS}.

A crucial tool in \cite{CKM} as well as in our work is positivity
of the Lyapunov exponent. To define it in the continuous case, let
$g_{\lambda}(n, \omega)$ denote the transfer matrix from $n-1/2$
to $n+1/2$ of
\begin{equation} \label{schrod}
-u^{\prime \prime}+ V_{\omega}u = \lambda u
\end{equation}
i.e. for any solution $u$ of $(\ref{schrod})$

\begin{equation} \label{transfermatrix}
\left( \begin{array}{c} u(n+1/2) \\ u^{\prime}(n+1/2) \end{array}
\right) = g_{\lambda}(n, \omega) \left( \begin{array}{c} u(n-1/2)
\\ u^{\prime}(n-1/2) \end{array} \right).
\end{equation}
For $n \in \N$, let
\begin{equation*}
U_{\lambda}(n,\omega) = g_{\lambda}(n,\omega) \cdot \ldots \cdot
g_{\lambda}(1,\omega).
\end{equation*}
The subadditive ergodic theorem (e.g. \cite{Carmona/Lacroix})
guarantees the existence of the Lyapunov exponent, i.e. the limit
\begin{equation*}
\gamma( \lambda) = \lim_{n \to \infty} \frac{1}{n} \E \left( \M
U_{\lambda}(n, \omega) \M \right).
\end{equation*}

Kotani Theory implies that $\gamma( \lambda)$ is positive for
almost every $\lambda \in \R$ since $H_{\omega}$ is
non-deterministic in the sense of Kotani, e.g.\ \cite{Kotani}. But
this is insufficient for a proof of localization, due to the fact
that the singular distribution of $\mu$ prevents spectral
averaging techniques from being used to prove absence of singular
continuous spectrum. For the discrete model $(\ref{discreteam})$,
one can instead use Furstenberg's Theorem on products of
independent random matrices to show that the Lyapunov exponent is
positive for all energies, which is the starting point of the
approach in \cite{CKM}.

We will use the same idea, but there is an additional, significant
difficulty to overcome. For the continuous model $(
\ref{contam})$, $( \ref{apot})$ there may exist a set of critical
energies at which Furstenberg's Theorem does not apply, and, in
most cases, $\gamma( \lambda)$ vanishes. This gives rise to the
exceptional set $M$ in Theorem $\ref{dynloc}$ and is a continuum
analog of the critical energies observed in \cite{Bi/Ger2} for the
random dimer model.

We manage to prove discreteness of the set of critical energies
$M$ for arbitrary single site potentials $f$ with support in
$[-1/2,1/2]$ and arbitrary periodic background $V_{{\rm per}}$ by using
methods from scattering theory. More precisely, we will consider
scattering at $f$ with respect to the periodic background
$V_{{\rm per}}$. For energies $\lambda$ in the interior of spectral
bands of the periodic operator $H_0 = -d^2/dx^2 + V_{{\rm per}}$, we
will introduce the reflection and transmission coefficients
$b(\lambda)$ and $a(\lambda)$ relative to $V_{{\rm per}}$. The roots of
$b(\lambda)$ give rise to critical energies where the Lyapunov
exponent vanishes and extended states exist. Since $f\not= 0$ it
can be shown that this set of energies is discrete. Other discrete
exceptional sets will be included in $M$, among them the band
edges of $H_0$ and roots of the analytic continuations of
$b(\lambda)$ and $a(\lambda)$ to the spectral gaps of $H_0$. Away
from $M$, positivity of $\gamma$ will follow from Furstenberg's
Theorem.

The fact that extended states exist at some critical energies does
not affect Theorem~\ref{exploc} since the discrete exceptional set
can not support continuous spectrum. It does, however, affect
dynamical localization in that the interval $I$ in
Theorem~\ref{dynloc} specifically excludes these critical values.

The existence of extended states at critical energies in certain
one-dimensional tight binding models or, equivalently, disordered
harmonic chains, is well known in the physics literature. Various
aspects of this are discussed in Section 10 of \cite{Li/Gr/Pa},
which also includes continuum models. Additional references can be
found through \cite{Datta/Kundu} and \cite{Bi/Ger2}. These works
also note the vanishing of single site reflection coefficients as
the basic mechanism for obtaining critical energies and discuss
the anomalous transport behavior near these energies. In several
of these models the discrete random potential is of the form
\begin{equation} \label{gendisand}
V_{\omega}(n) = \sum_{k\in\Z} q_k(\omega) f(n-k\ell),
\end{equation}
where the single site potential $f:\Z\to \R$ is supported on $\{0,
\ldots, \ell-1\}$. For example, in the dimer model as studied in
\cite{Bi/Ger2} one sets $\ell=2$, $f=\chi_{ \{ 0,1 \} }$. While in
our work we focus on a systematic study of exceptional energies in
continuum Anderson models, we expect that our methods can be
adapted to yield exponential and dynamical localization for
general discrete Anderson models of the form (\ref{gendisand}).
While we exclude neighborhoods of critical energies in our
discussion of dynamical localization, as does \cite{Bi/Ger2}, it
is an interesting open problem to describe the transport behavior
near critical energies rigorously.

A review of results on positivity of the Lyapunov exponent which
have been obtained from Furstenberg's Theorem is given in
Section~14.A of \cite{Figotin/Pastur}. Most of these results were
obtained for discrete Schr\"odinger operators and Jacobi matrices,
see e.g.\ \cite{Ishii/Matsuda, Ishii, Figotin}. Results for
continuum Anderson models (\ref{contam}), (\ref{apot}) had been
restricted to special exactly solvable cases which could be
reduced to discrete models. For example, the set where $\gamma$
vanishes was characterized for the cases where $V_{{\rm per}}=0$ and
either $f$ is a $\delta$-point-interaction \cite{Ishii} or $f=
\chi_{[-1/2,1/2]}$ \cite{Benderskii/Pastur}. For the latter, if it
is assumed that supp$(\mu) = \{0,\xi\}$ for some $\xi>0$, then one
gets the critical energies $\{\lambda: \gamma(\lambda)=0\} = \{
\xi+\pi^2 n^2, n\in\Z\}$.

The first paper where reflection coefficients are used rigorously
in combination with Furstenberg's Theorem to characterize the
exceptional set for general $f$ is \cite{Kostrykin/Schrader}. Many
of the ideas used in Section~2 below can be found there. They
assume $V_{{\rm per}}=0$ and thus can work with the ``classical''
reflection and transmission coefficients for scattering at $f$.
Also, \cite{Kostrykin/Schrader} assumes that the distribution
$\mu$ of $q_n$ has continuous density, but the methods used can be
modified to include some examples with singular distributions. We
also mention the recent paper \cite{Kostrykin/Schrader2} which
contains bounds for the Lyapunov exponent, which might be useful
to study transport properties near critical energies.

That energies with vanishing reflection coefficient at a single
site have zero Lyapunov exponent was also observed for random
displacement models in \cite{Sims/Stolz}. That the latter model
shows similar properties as the Bernoulli-Anderson model is not
surprising: Both models display identical single site potentials
at random distances.

To our knowledge, the only previous work on localization for {\it
continuum} Bernoulli-Anderson models is \cite{Carmona1}. This
work, in combination with remarks in \cite{CLN}, looks at the
example $V_{{\rm per}}=0$, $f=\chi_{[-1/2,1/2]}$, supp$(\mu) = \{0,1\}$,
and states exponential localization for all energies.
Unfortunately, it seems that some details of the proof, which is
sketched in \cite{Carmona1} and quite different from ours, have
never appeared in print.

We now outline the contents of the remaining sections. In
Section~2 we prove positivity of the Lyapunov exponent away from a
discrete set using Furstenberg's Theorem. A key tool is to show
that the identical vanishing of $b(\lambda)$ on a band of $H_0$
implies that $f=0$. This generalizes the well known fact that
there are no compactly supported solitons to scattering at
periodic background. Sections 3 and 4 establish that the Lyapunov
exponent and the integrated density of states are H\"older
continuous away from the exceptional set. Both sections rely
heavily on known results for the discrete case, which can all be
found in \cite{Carmona/Lacroix}. We outline these results and
discuss the necessary changes for continuum models. In particular,
we make use of a version of the Thouless formula by Kotani
\cite{Kotani}.

The next two sections provide the main ingredients for the start
of a multiscale analysis. In Section~5 a Wegner estimate is
proven, again away from the exceptional set. Here we can closely
follow the argument of \cite{CKM}. Section~6 provides an initial
length scale estimate (ILSE) for the same energies. Here our
argument differs somewhat from the approach in \cite{CKM}. While
the latter uses both, positivity of the Lyapunov exponent and the
Wegner estimate, to get an ILSE, we prefer to use an approach
which shows that ILSE follows directly from positivity of
$\gamma$. This uses some results on large deviations for $\gamma$,
which can all be found in \cite{Bougerol/Lacroix}. We note that
this also gives a new, more natural proof of an ILSE for discrete
models. Having established all what is necessary for a multiscale
analysis, we can then prove Theorems~\ref{exploc} and \ref{dynloc}
in Section~7 by merely referring to well known results, e.g.\
\cite{Stollmann} and \cite{DS}.

Appendix A contains some a priori estimates for solutions of the
Schr\"odinger equation which we use frequently. In Appendix B some
basic facts about cocycles and the existence and uniqueness of
invariant measures for group actions are listed. They are used in
Section~3.

\vspace{.3cm}

\noindent {\bf Acknowledgement:} We are grateful to B.\ Simon for
pointing us to the simple proof of Lemma \ref{zero} below, i.e.
that there are no compactly supported ``solitons" for scattering
at periodic background. After this we learned a different proof of
this fact from E. Korotyaev. We also acknowledge useful
discussions with L.\ Pastur and R.\ Schrader.

\setcounter{equation}{0}


\section{Positivity of the Lyapunov exponent}\label{lyapunovsec}

\subsection{The Floquet Solutions}
We start by collecting some facts from Floquet theory for the
periodic operator $H_0:= - \frac{d^2}{dx^2} + V_{{\rm per}}$, see e.g.
\cite{Eastham}. For any $z \in \C$, let $u_N( \cdot, z)$ and $u_D(
\cdot, z)$ denote the solutions of
\begin{equation} \label{pereqn}
-u^{\prime \prime} + V_{{\rm per}}u = zu
\end{equation}
with $u_N(-1/2)=u^{\prime}_D(-1/2)=1$ and
$u_N^{\prime}(-1/2)=u_D(-1/2)=0$. The transfer matrix of
$(\ref{pereqn})$ from $-1/2$ to $1/2$ is the matrix
\begin{equation*}
g_0(z) = \left( \begin{array}{cc} u_N(1/2,z) & u_D(1/2,z) \\
u^{\prime}_N(1/2,z) & u^{\prime}_D(1/2,z) \end{array} \right),
\end{equation*}
which is entire in $z$ as solutions of $(\ref{pereqn})$ (resp.
their derivatives) are, for each fixed $x$, entire in $z$. The
eigenvalues of $g_0(z)$ are the roots of
\begin{equation} \label{evaleqn}
\rho^2 - D(z) \rho +1 =0,
\end{equation}
i.e.,
\begin{equation} \label{rhoeqn}
\rho_{\pm}(z) = \frac{D(z) \pm \sqrt{D(z)^2 -4}}{2},
\end{equation}
where $D(z)= \mbox{Tr}[g_0(z)]$. As roots of $(\ref{rhoeqn})$, the
functions $\rho_{\pm}$ are algebraic with singularities at points
with $D(z)= \pm 2$.

The spectrum of $H_0$, $\sigma(H_0)$, consists of bands which are
given by the sets of real energies $\lambda$ for which $D(
\lambda) \leq 2$. Let $(a,b)$ be a stability interval of $H_0$;
i.e. a maximal interval such that $D(\lambda) < 2$ for every
$\lambda \in (a,b)$. As both $u_N( \cdot, \lambda)$ and $u_D(
\cdot,\lambda)$ are real for $\lambda \in (a,b)$, one has that
\begin{equation*}
\rho_{\pm}(\lambda) = \frac{1}{2} \left( D(\lambda) \pm i \sqrt{4-
D(\lambda)^2} \right),
\end{equation*}
$|\rho_{\pm}(\lambda)|=1$, and $\rho_-(\lambda) = \overline{
\rho_+(\lambda)}$. Let
\begin{equation*}
S:= \{ z \in \C : z = \lambda + i \eta \ \ \mbox{where} \ \ a
< \lambda < b \ \ \mbox {and} \ \ \eta \in \R \}
\end{equation*}
be the vertical strip in the complex plane containing $(a,b)$. For
$z=\lambda+i\eta \in S$ one has that each of the following are
equivalent: (i) $|\rho_{\pm}(z)|$ =1, (ii) $\eta $ = 0, and (iii)
$D(z) \in (-2,2)$. That $(ii) \Rightarrow (iii)$ and $(iii)
\Rightarrow (i)$ are clear. To see that $(i) \Rightarrow (ii)$
assume that $(i)$ is true for some $z=\lambda+i \eta$ where $\eta
\neq 0$. As both $\rho_{\pm}$ have modulus 1, then all solutions
of $(\ref{pereqn})$ are bounded. By Weyl's alternative, however,
if $\eta \neq 0$, then there exists a solution in $L^2$ near $+
\infty$, the Weyl solution, while all other solutions are
unbounded. This is a contradiction.

The arguments above imply that $\rho_{\pm}$ have analytic
continuations to all of $S$, and that the only possible algebraic
singularities occur at $a$ and $b$. In addition, as $\rho_+(z)
\rho_-(z) = \mbox{det}[g_0(z)] = W[u_N,u_D]=1$, we have by
continuity of $| \cdot|$ that exactly one of $\rho_+$ and $\rho_-$
satisfies $| \rho(z)|<1$ in the upper half plane. Without loss of
generality, let us denote by $\rho_+$ the eigenvalue for which
$|\rho_+(\lambda+i\eta)|<1$ for all $\eta
>0$ and $\lambda\in (a,b)$. This corresponds to choosing a
branch of the square root in $( \ref{rhoeqn})$. Then, $\rho_-$
satisfies $|\rho_-(\lambda+i\eta)|>1$ for all $\eta >0$ and
$\lambda\in (a,b)$. Since we also have $|\rho_{\pm}( \lambda)|=1$
for $\lambda \in (a,b)$, it follows from the Schwarz reflection
principle (apply a linear fractional transformation) that
$|\rho_+(\lambda+i\eta)|>1$ and $|\rho_-(\lambda+i\eta)|<1$ for
all $\eta <0$ and $\lambda\in (a,b)$.

For $z \in S$, let $v_{\pm}(z)$ be the eigenvectors of $g_0(z)$
corresponding to $\rho_{\pm}(z)$ with first component normalized
to be one; i.e.
\begin{equation} \label{evec}
v_{\pm}(z) = \left( \begin{array}{c} 1 \\ c_{\pm}(z) \end{array}
\right).
\end{equation}
One may easily calculate that
\begin{equation} \label{cpm}
c_{\pm}(z) = \frac{ \rho_{\pm}(z) - u_N(1/2,z)}{u_D(1/2,z)}.
\end{equation}

As $u_D(1/2,z)$ is never zero in $S$ (Dirichlet eigenvalues of
$H_0$ are either in the gaps of $\sigma(H_0)$ or at the band
edges), $v_{\pm}$ are analytic in $S$. In particular, as
$u_D^{-1}$ has at most a pole at $a$ and $b$, then $c_{\pm}$, and
hence $v_{\pm}$, have at worst algebraic singularities at $a$ and
$b$. (More precisely, $c_{\pm}$ are branches of a multi-valued
analytic function with algebraic singularities at the boundaries
of stability intervals.) Let $\phi_{\pm}( \cdot,z)$ be the Floquet
solutions of $(\ref{pereqn})$; i.e. the solutions satisfying
\begin{equation} \label{fbcs}
\left( \begin{array}{c} \phi_{\pm}(-1/2,z) \\
\phi^{\prime}_{\pm}(-1/2,z) \end{array} \right) = v_{\pm}(z).
\end{equation}

We first note that $\phi_{\pm}( \cdot, \lambda+ i \eta) \in L^2$
near $\pm \infty$ if $\eta >0$, and $\phi_{\pm}( \cdot, \lambda+ i
\eta) \in L^2$ near $\mp \infty$ if $\eta <0$. Thus in this
setting the Floquet solutions are the Weyl solutions. Secondly,
for fixed $x$, $\phi_{\pm}(x, \cdot)$ are analytic in $S$ with at
worst algebraic singularities at $a$ and $b$ arising from the
singularities in the initial conditions $v_{\pm}$. Lastly, $\{
\phi_+( \cdot,z), \phi_-(\cdot,z) \}$ are a fundamental system of
$(\ref{pereqn})$ for every $z \in S$, as $\rho_+(z) \neq
\rho_-(z)$ on $S$.

\subsection{Scattering with respect to a periodic background}

Let $0 \neq f \in L^1_{loc}(\R)$ be real valued with supp$(f)
\subset [-1/2,1/2]$. Consider the operator
\begin{equation*}
H = - \frac{d^2}{dx^2} +V_{{\rm per}} + f
\end{equation*}
in $L^2( \R)$. Take $z \in S$ and let $u_+$ be the solution of
\begin{equation} \label{perteqn}
-u^{\prime \prime} + (V_{{\rm per}} +f)u =zu
\end{equation}
satisfying
\begin{equation} \label{+jost}
u_+(x,z) = \left\{ \begin{array}{cc} \phi_+(x,z) & \mbox{for } x
\leq -1/2 \\ a(z) \phi_+(x,z) + b(z) \phi_-(x,z) & \mbox{for } x
\geq 1/2. \end{array} \right.
\end{equation}
Since $\phi_{\pm}$ are linearly independent for $z \in S$, this
defines $a(z)$ and $b(z)$ uniquely. If $V_{{\rm per}}=0$, then $a$ and
$b$ are related to the classical transmission and reflection
coefficients by $a=t^{-1}$ and $b=rt^{-1}$. In particular,
vanishing of $b$ is equivalent to vanishing reflection
coefficient. Thus $b$ and $u_+$ take the role of a (modified)
reflection coefficient and Jost solution relative to the periodic
background $V_{{\rm per}}$.

Since for $\lambda \in (a,b)$ we know $\phi_-(x,\lambda)=
\overline{\phi_+(x,\lambda)}$ from $(\ref{evec})$, $(\ref{cpm})$,
and $(\ref{fbcs})$, by taking $u_-$ to be the solution of
$(\ref{perteqn})$ with $u_-(x,\lambda)= \phi_-(x,\lambda)$ for $x
\leq -1/2$ we get that for $x \geq 1/2$:
\begin{equation} \label{-jost}
u_-(x,\lambda) = \overline{a(\lambda)} \phi_-(x,\lambda) +
\overline{b(\lambda)} \phi_+(x,\lambda).
\end{equation}
Using constancy of the non-zero Wronskian, we arrive at the
familiar relation
\begin{equation} \label{abeqn}
|a(\lambda)|^2 - |b(\lambda)|^2 = 1,
\end{equation}
for $\lambda \in (a,b)$, corresponding to $|r|^2+|t|^2=1$.
\begin{Proposition} $a( \cdot)$ and $b( \cdot)$,
defined on $S$ as above, are branches of multi-valued analytic
functions with at most algebraic singularities at boundaries of
stability intervals.
\end{Proposition}

\begin{proof} Recall that $u_+$ is the solution of
$(\ref{perteqn})$ with
\begin{equation*}
\left( \begin{array}{c} u_+(-1/2,z) \\ u^{\prime}_+(-1/2,z)
\end{array} \right) = \left(  \begin{array}{c} \phi_+(-1/2,z) \\
\phi^{\prime}_+(-1/2,z) \end{array} \right) = v_+(z).
\end{equation*}
Thus $\left( u_+(1/2,z),u^{\prime}_+(1/2,z) \right)^t$ is analytic
in $S$ with at worst algebraic singularities at $a$ and $b$, i.e.
the boundaries of the stability interval. But, as was determined
before, both $\left( \phi_{\pm}(1/2,z),\phi_{\pm}^{\prime}(1/2,z)
\right)^t$ have identical analytic properties, and by the
definition of $a(z)$ and $b(z)$,
\begin{equation} \label{abvec}
\left( \begin{array}{c} a(z) \\ b(z)  \end{array} \right) = \left(
\begin{array}{cc} \phi_+(1/2,z) & \phi_-(1/2,z) \\ \phi^{\prime}_+(1/2,z) & \phi^{\prime}_-(1/2,z)
\end{array} \right)^{-1} \left( \begin{array}{c} u_+(1/2,z) \\
u^{\prime}_+(1/2,z)
\end{array} \right),
\end{equation}
so we are done.
\end{proof}

\begin{Lemma} \label{zero} If $b(\lambda)=0$ for all $\lambda \in (a,b)$, then $f$ is
identically zero.
\end{Lemma}
\begin{proof}
Suppose that $b(\lambda)=0$ for all $\lambda \in (a,b)$, and hence
all $z \in S$ by analyticity. We know then that for every $\lambda
\in (a,b)$ and $\eta>0$, the Jost solution $u_+(x,\lambda+i \eta)$
is $a(\lambda + i \eta) \phi_+(x,\lambda+i\eta)$ (for all $x
\geq1/2$); i.e. $u_+$ is exponentially decaying in this region of
the upper half plane. Thus $u_+$ is the Weyl solution for the
perturbed equation $(\ref{perteqn})$. We may therefore calculate
the Weyl-Titchmarsh $M$-function (see e.g. \cite{Codd/Lev}),
$M_{V_{{\rm per}}+f}$, for $(\ref{perteqn})$ on the half line $(-1/2,
\infty)$,
\begin{equation*}
M_{V_{{\rm per}}+f}(\lambda+i \eta) = \frac{u^{\prime}_+(-1/2,\lambda+i
\eta)}{u_+(-1/2,\lambda+i \eta)} =
\frac{\phi^{\prime}_+(-1/2,\lambda+i \eta)}{\phi_+(-1/2,\lambda+i
\eta)} = M_{V_{{\rm per}}}(\lambda+i\eta),
\end{equation*}
where the latter is the $M$-function of $(\ref{pereqn})$ on
$(-1/2, \infty)$. As the $M$-functions are analytic in the entire
upper half plane, we conclude that
$M_{V_{{\rm per}}+f}(z)=M_{V_{{\rm per}}}(z)$ for all $z \in \C^+$. Thus, by
the recent results of Simon \cite{Simon1} and Gesztesy and Simon
\cite{Gesztesy/Simon}, which provide generalizations (applicable
in our setting) to the classical results of
\cite{Borg,Marchenko1,Marchenko2}, this implies $f$ is identically
zero.
\end{proof}

As we assume throughout that $f$ is not identically zero, then we
know that $\{ \lambda \in (a,b): b(\lambda)=0 \}$ is in fact
finite as accumulations at algebraic singularities are not
possible.

Now we consider a gap. Take $\alpha$ such that $- \infty \leq
\alpha < a <b$ and $( \alpha, a)$ is a maximal, non-trivial gap in
the spectrum of $H_0$ (if $a= \inf{ \sigma(H_0)}$, then $\alpha =
- \infty$). Consider the following split strip:
\begin{equation*}
S^{\prime}:= \{ z=\lambda+i\eta: \alpha < \lambda < b, \eta \in \R
\} \setminus [ \mbox{a},\mbox{b}).
\end{equation*}
For $i=1,2$, let $\rho_i(z)$ be the branches of $(\ref{evaleqn})$
with $|\rho_1(z)|<1$ and $|\rho_2(z)|>1$ for all $z \in
S^{\prime}$, which are well defined since $[a,b)$ is excluded. We
first note that $\rho_1 = \rho_+$ on the upper half of $S$, but
$\rho_1= \rho_-$ on the lower half of $S$. Secondly, as before, it
can be seen that $\rho_i$ are analytic in $S^{\prime}$ and have at
worst algebraic singularities at $\alpha$, $a$, and $b$, and
therefore they may be continued analytically across $(a,b)$. In
particular, $\rho_i$ is the analytic continuation of $\rho_j$,
where $i,j \in \{1,2\}$ with $i \neq j$.

For $z \in S^{\prime}$, choose analytic eigenvectors $v_i(z)$ of
$g_0(z)$ to corresponding $\rho_i(z)$, see Ch.II.4 \cite{Kato}.
Taking $\phi_i$ to be the solutions of $(\ref{pereqn})$ with $(
\phi_i(-1/2,z), \phi^{\prime}_i(-1/2,z) )^t=v_i(z)$ for $z \in
S^{\prime}$, we see again that $\phi_1( \cdot, z)$ (resp.,
$\phi_2( \cdot,z)$) is in $L^2$ near $+ \infty$ (resp., $-
\infty$); i.e. they are the Weyl solutions. Set $u_i$ to be the
Jost solutions of $(\ref{perteqn})$ satisfying
\begin{equation} \label{jostsols}
u_i( \cdot, z) = \left\{ \begin{array}{cc} \phi_i(x, z) & x \leq
-1/2
\\ a_i(z) \phi_i(x,z) + b_i(z)\phi_j(x,z) & x \geq 1/2 \end{array}
\right.
\end{equation}
for $z \in S^{\prime}$ and the same $i,j$ convention used above.
As in $(\ref{abvec})$ above, one sees that $a_1(z)$ and $b_1(z)$
are analytic in $S^{\prime}$. In fact, in the upper half $S_+$ of
$S$, they coincide with $a(z)$ and $b(z)$, since for $z \in S_+$,
we have that $v_+(z)$ and $v_1(z)$ coincide up to a constant
multiple, which cancels in $(\ref{abvec})$ and the corresponding
expression for  $a_1(z)$ and $b_1(z)$. Thus  $a_1(z)$ and $b_1(z)$
are analytic continuations of the restrictions of  $a(z)$ and
$b(z)$ to $S_+$. Similarly, it is seen that $a_2(z)$ and $b_2(z)$
are analytic continuations of the restrictions of  $a(z)$ and
$b(z)$ to the lower half $S_-$ of $S$. Thus, $a_1(z), b_1(z),
a_2(z)$, and $b_2(z)$ have at most algebraic singularities at
$\alpha$, $a$, and $b$. In particular, since they cannot vanish
identically by $f \neq 0$, the set
\begin{equation} \label{abroots}
\{ \lambda \in (\alpha,b):
a_1(\lambda)b_1(\lambda)a_2(\lambda)b_2(\lambda)=0 \}
\end{equation}
is discrete and, if $\alpha \neq - \infty$, finite.

\subsection{Transfer matrices and positivity of $\gamma$}
To prove positivity of $\gamma(\lambda)$, we will investigate
properties of the transfer matrices. As our minimal assumption is
that the support of the distribution contains the points 0 and 1,
we know that there are at least two non-trivial transfer matrices:
the free matrix $g_0(\lambda)$ corresponding to $(\ref{pereqn})$
and the perturbed matrix $g_1(\lambda)$ corresponding to
$(\ref{perteqn})$. Set $G(\lambda)$ to be the closed subgroup of
${\rm SL}(2, \R)$ generated by $\{g_q(\lambda): q \in \mbox{supp}( \mu)
\}$. Let $P( \R^2)$ be the projective space, i.e. the set of the
directions in $\R^2$ and $\overline{v}$ be the direction of $v \in
\R^2 \setminus \{ 0 \}$. Note that ${\rm SL}(2, \R)$ acts on $P( \R^2)$
by $g \overline{v} = \overline{gv}$. We say that $G \subset SL(2,
\R)$ is strongly irreducible if and only if there is no finite
$G$-invariant set in $P( \R^2)$.

\begin{Theorem} \label{gamma+} Given a family of operators
$\{ H_{\omega} \}$, as in $(\ref{contam})$, $(\ref{apot})$, with $f \neq
0$ and $0,1 \in \mbox{supp}( \mu)$, then there exists a discrete
set $M \subset \R$ such that $G( \lambda)$ is non-compact and
strongly irreducible for all $\lambda \in \R \setminus M$. In
particular, $\gamma(\lambda)>0$ for all $\lambda \in \R \setminus
M$.
\end{Theorem}

We will first prove that $G(\lambda)$ is not compact by showing
that a sequence of elements has unbounded norm. This argument will
be valid for all but a discrete set of $\lambda$'s. Once
non-compact, then the group $G$ is known to be strongly
irreducible if and only if for each $\overline{v} \in P(\R^2)$,
\begin{equation} \label{3dir}
\#\{g \overline{v}: g \in G\} \geq 3,
\end{equation}
see \cite{Bougerol/Lacroix}. We shall prove that this condition is also
satisfied for all but a slightly larger, yet still discrete set of
$\lambda$'s. The conclusion concerning positivity of $\gamma$
follows from a general Theorem of Furstenberg which, in this
context, states that if the group $G(\lambda)$ is non-compact and
strongly irreducible, then $\gamma(\lambda) >0$; see also
\cite{Bougerol/Lacroix}. Note that non-compactness of $G(
\lambda)$ and $(\ref{3dir})$ are both properties which are
preserved if supp$(\mu)$ is increased. Thus we will assume
w.l.o.g. that supp$(\mu)= \{0,1\}$, i.e. $G(\lambda)$ is generated
by $g_0( \lambda)$ and $g_1( \lambda)$.

In the construction of $M$ we will treat stability intervals and
spectral gaps of $H_0$ separately. We will first show that in any
given stability interval $(a,b)$, there is at most a finite number
of $\lambda$'s such that Furstenberg's Theorem does not apply. We
then show the same for finite spectral gaps $(\alpha,a)$ of $H_0$.
For the spectral gap $(- \infty$, inf $\sigma(H_0)$), we allow for
accumulation of exceptional energies at $-\infty$. Finally, we
join all these exceptional sets and the endpoints of stability
intervals to get the discrete set $M$.

\begin{Lemma} \label{conj} Let $\lambda$ be in a stability interval $(a,b)$ of
$H_0$. Then,
\begin{equation} \label{greducf}
g_1(\lambda) = C(\lambda)
\tilde{g}_0(\lambda)s(\lambda)C(\lambda)^{-1},
\end{equation}
where
\begin{equation} \label{cg}
C(\lambda) = \left( \begin{array}{cc} 1 & 0 \\
{\rm Re}[c(\lambda)] & {\rm Im}[c(\lambda)] \end{array} \right),
\ \ \tilde{g}_0(\lambda) = \left( \begin{array}{cc}
{\rm Re}[\rho(\lambda)] & {\rm Im}[\rho(\lambda)]
\\ -{\rm Im}[\rho(\lambda)] & {\rm Re}[\rho(\lambda)]
\end{array} \right),
\end{equation}
\begin{equation} \label{scat}
s(\lambda) = \left( \begin{array}{cc}
{\rm Re}[a(\lambda)+b(\lambda)] &
{\rm Im}[a(\lambda)+b(\lambda)]
\\ -{\rm Im}[a(\lambda)-b(\lambda)]
& {\rm Re}[a(\lambda)-b(\lambda)] \end{array} \right),
\end{equation}
and $c(\lambda) =c_+(\lambda) = \overline{c_-(\lambda)}$,
$\rho(\lambda) = \rho_+(\lambda) = \overline{ \rho_-(\lambda)}$ as
given in $(\ref{rhoeqn})$ and $(\ref{cpm})$.
\end{Lemma}
\begin{proof} Since $\lambda \in (a,b)$, we have that $c_+(\lambda) =
\overline{c_-(\lambda)}$, $\rho_+(\lambda) = \overline{
\rho_-(\lambda)}$, and ${\rm Im}[c(\lambda)] \neq 0$ from
$(\ref{rhoeqn})$ and $(\ref{cpm})$. Thus $C(\lambda)$ is
invertible. For the rest of the proof we will drop the fixed
parameter $\lambda$. First note that expressing the solutions
$u_N$ and $u_D$ in terms of the Jost solutions $u_{\pm}$ yields
\begin{equation*}
g_1(\lambda) = \left( \begin{array}{cc} u_+(1/2) & u_-(1/2) \\
u^{\prime}_+(1/2) & u^{\prime}_-(1/2) \end{array} \right) \left(
\begin{array}{cc} 1 & 1 \\ c &  \overline{c} \end{array}
\right)^{-1}.
\end{equation*}
Clearly,
\begin{equation*}
\left( \begin{array}{cc} u_+(1/2) & u_-(1/2) \\ u^{\prime}_+(1/2)
& u^{\prime}_-(1/2) \end{array} \right) = \left( \begin{array}{cc}
\phi_+(1/2) & \phi_-(1/2) \\ \phi^{\prime}_+(1/2) &
\phi^{\prime}_-(1/2)
\end{array} \right) \left( \begin{array}{cc} a & \overline{b} \\ b &
\overline{a} \end{array} \right),
\end{equation*}
and using that $\phi_{\pm}$ are the Floquet solutions we have
\begin{equation*}
\left( \begin{array}{cc} \phi_+(1/2) & \phi_-(1/2) \\
\phi^{\prime}_+(1/2) & \phi^{\prime}_-(1/2)
\end{array} \right) =  \left( \begin{array}{cc} 1 & 1 \\ c & \overline{c} \end{array}
\right)  \left( \begin{array}{cc} \rho & 0 \\ 0 & \overline{\rho}
\end{array} \right).
\end{equation*}
Hence we see
\begin{equation} \label{greduci}
g_1(\lambda) = \left( \begin{array}{cc} 1 & 1 \\ c & \overline{c}
\end{array} \right) \left( \begin{array}{cc} \rho & 0 \\ 0 &
\overline{\rho}
\end{array} \right) \left( \begin{array}{cc} a & \overline{b} \\ b &
\overline{a} \end{array} \right) \left( \begin{array}{cc} 1 & 1 \\
c & \overline{c} \end{array} \right)^{-1}.
\end{equation}
Let
\begin{equation*}
Q:= \frac{1}{2}  \left( \begin{array}{cc} 1 & -i \\ 1 & i
\end{array} \right),
\end{equation*}
then it is easily checked that
\begin{equation*}
Q^{-1} \left( \begin{array}{cc} a & \overline{b} \\ b &
\overline{a} \end{array} \right) Q = s, \ \  Q^{-1} \left(
\begin{array}{cc} \rho & 0 \\ 0 & \overline{\rho}
\end{array} \right) Q = \tilde{g}_0, \mbox{ and }
 \left( \begin{array}{cc} 1 & 1 \\
c & \overline{c} \end{array} \right)Q=C,
\end{equation*}
with $C$, $\tilde{g}_0$ and $s$ as defined in $(\ref{cg})$,
$(\ref{scat})$. Thus $(\ref{greducf})$ follows from
$(\ref{greduci})$.
\end{proof}

Since $g_0(\lambda)$ corresponds to trivial scattering, it follows
from $(\ref{greducf})$ that
\begin{equation} \label{freduc}
g_0(\lambda) = C(\lambda) \tilde{g}_0(\lambda)C(\lambda)^{-1}.
\end{equation}
Let $\tilde{G}(\lambda)$ be the subgroup of SL$(2, \R)$ generated
by $\tilde{g}_0(\lambda)$ and $\tilde{g}_0(\lambda)s(\lambda)$ or
equivalently, by $\tilde{g}_0(\lambda)$ and $s(\lambda)$.
$\tilde{G}(\lambda)$ is conjugate to $G(\lambda)$, and thus
$G(\lambda)$ is non-compact and strongly irreducible if and only
if $\tilde{G}(\lambda)$ is non-compact and strongly irreducible.

Before proving Theorem $\ref{gamma+}$, we note that for every
$\lambda$ in a stability interval $(a,b)$, $\rho( \lambda) = e^{i
\omega}$, where $\omega = \omega( \lambda) \in (0, \pi)$. Thus,
\begin{equation} \label{free}
\tilde{g}_0(\lambda) = \left( \begin{array}{cc} \cos( \omega) &
\sin( \omega) \\ - \sin( \omega) & \cos( \omega) \end{array}
\right)
\end{equation}
is merely a rotation. In particular, for any $\lambda \in (a,b)$
satisfying $b( \lambda)=0$, one has that $a( \lambda) = e^{i
\alpha}$, by $(\ref{abeqn})$, and therefore $s( \lambda)$ is
reduced to a rotation matrix as well. For these finitely many
$\lambda$'s (see the paragraph immediately following the proof of
Lemma $\ref{zero}$), $\tilde{G}(\lambda)$ is a group of rotations
and thereby compact. Moreover, by $(\ref{greducf})$ and
$(\ref{freduc})$ we know that at such energies, the norm of
products of the transfer matrices does not grow: hence $\gamma(
\lambda)=0$. Let $ \tilde{M}^{(a,b)}
:= \{ \lambda \in (a,b): b( \lambda) =0 \}$. It is clear then
that only away from this set, $\tilde{M}^{(a,b)}$, can we hope to
apply Furstenberg's Theorem and prove positivity of $\gamma$.

\medskip

\noindent {\it Proof of Theorem $\ref{gamma+}$.} {\it The
stability intervals:} We first consider energies in a stability
interval $(a,b)$. Let $\lambda \in (a,b) \setminus
\tilde{M}^{(a,b)}$, i.e. such that $b( \lambda) \neq 0$. Set $a(
\lambda) = Ae^{i \alpha}$, $b( \lambda) = Be^{ i \beta}$ and see
that $( \ref{abeqn})$ implies $A^2-B^2=1$ with $A>B>0$, $A+B>1$,
and $A-B<1$. In this notation one may recalculate
\begin{equation*}
s(\lambda) = A \left( \begin{array}{cc} \cos( \alpha) & \sin(
\alpha) \\ - \sin( \alpha) & \cos(\alpha) \end{array} \right) +B
\left( \begin{array}{cc} \cos( \beta) & \sin(\beta)
\\ \sin(\beta) & -\cos(\beta) \end{array} \right).
\end{equation*}

We wish to show that a sequence of elements in
$\tilde{G}(\lambda)$ has unbounded norm. To do so, consider an
arbitrary element of $P( \R^2)$, represented by
\begin{equation*}
v( \theta) := \left( \begin{array}{c} \cos(\theta)
\\ \sin( \theta) \end{array} \right).
\end{equation*}
One sees that the relation
\begin{equation*}
s(\lambda)v( \theta) = A v( \theta - \alpha) + B v( \beta -
\theta)
\end{equation*}
suggests the choice $\theta^{\prime}:= \frac{1}{2}( \alpha +
\beta)$ yielding
\begin{equation*}
s(\lambda)v( \theta^{\prime})= (A+B)v([\beta -\alpha]/2).
\end{equation*}
Defining $R( \theta):= \M s(\lambda)v( \theta) \M^2$, a short
calculation, using $A^2-B^2=1$, shows that
\begin{equation*}
R( \theta)-1 =2B \left[ B+A \cos(\alpha + \beta - 2\theta)
\right].
\end{equation*}
As a consequence, we see that $R( \theta)-1$ is $\pi$-periodic and
has exactly two roots in $[0,\pi)$. In particular, $R( \theta)=1$
if and only if $\cos( \alpha + \beta - 2 \theta) = - \frac{B}{A}$,
which shows that the distance between the zeros of $R( \theta)-1$
is not equal to $\frac{ \pi}{2}$: recall $B \neq 0$ (and $B \neq
A$). Similarly, $R( \theta)>1$ if and only if $\cos( \alpha +
\beta - 2 \theta) > - \frac{B}{A}$ and hence $| \{ \theta \in [0,
\pi): R( \theta)>1 \}| > \frac{ \pi}{2}$. As a result, there
exists a compact interval $K$ (not necessarily in $[0, \pi)$) with
$|K|> \frac{\pi}{2}$ and $\M s(\lambda)v( \theta) \M > c
> 1$ for all $\theta \in K$.

Now, applying $s(\lambda)$ once to $v( \theta^{\prime})$,
$\theta^{\prime}$ as above, produces a new vector with norm
greater than one, but the direction, initially $\theta^{\prime}$,
is possibly altered. A vector with this new direction may not
increase in norm by directly applying $s(\lambda)$ again. We may,
however, apply the modified free matrix, $(\ref{free})$, and
rotate this new direction by $\omega$. As $\omega \in (0, \pi)$,
then finitely many applications of $\tilde{g}_0(\lambda)$ produces
a vector $v$ with direction in $K$. Once in $K$, a direct
application of $s(\lambda)$ does increase the norm size uniformly
by $c>1$, as indicated above. In this manner, we can produce a
sequence of elements in $\tilde{G}(\lambda)$ with unbounded norm.

To finish the proof of Theorem $\ref{gamma+}$ in the stability
interval $(a,b)$, we choose
\begin{equation*}
M^{(a,b)}:= \{ \lambda \in (a,b): \lambda \in \tilde{M}^{(a,b)}
\mbox{ or } D(\lambda) = 0 \}.
\end{equation*}
Clearly, $M^{(a,b)}$ is finite, and by the above
$\tilde{G}(\lambda)$ is not compact for $\lambda \in (a,b)
\setminus M^{(a,b)}$. It remains the check ($\ref{3dir}$) for
these values of $\lambda$. But this is trivial as $D(\lambda) \neq
0$ implies $\omega \neq \frac{\pi}{2}$, and therefore the free
transfer matrix produces three distinct elements in projective
space.

\noindent {\it The spectral gaps:} For all energies $\lambda \in
(\alpha, a)$, a maximal spectral gap of $H_0$, $G(\lambda)$ is
non-compact. This is easily seen as $\lambda \in (\alpha,a)$
implies $|D(\lambda)|>2$ and hence the free transfer matrix,
$g_0$, has an eigenvalue $|\rho| = |\rho_2| > 1$. Thus repeated
iteration of $g_0$ on $\overline{v}_2$, the direction of $v_2$,
produces an unbounded sequence of elements. It remains to check
($\ref{3dir}$), which we will do for $\lambda \in (\alpha,a)
\setminus M^{(\alpha,a)}$, where
\begin{equation*}
M^{(\alpha,a)}:= \{ \lambda \in (\alpha,a):
a_1(\lambda)b_1(\lambda)a_2(\lambda)b_2(\lambda)=0 \}.
\end{equation*}
As was clear in $(\ref{abroots})$, $M^{(\alpha,a)}$ is discrete,
and finite if $\alpha \neq - \infty$.

Let $\lambda \in (\alpha,a) \setminus M^{(\alpha,a)}$, i.e.
$a_i(\lambda) \neq 0$ and $b_i(\lambda) \neq 0$ for $i=1,2$. By
($\ref{jostsols}$), this is equivalent to
\begin{equation} \label{unstable}
s(\lambda) \overline{v}_i \notin \{ \overline{v}_1, \overline{v}_2
\},
\end{equation}
where $ \overline{v}_i$ are the directions of the eigenvectors
$v_i$ of $g_0(\lambda)$, again for $i=1,2$. If $\overline{v}
\notin \{ \overline{v}_1, \overline{v}_2\}$, then $\# \{
g_0(\lambda)^n \overline{v}: n \in \Z \} = \infty$. If, on the
other hand, $\overline{v} \in \{ \overline{v}_1, \overline{v}_2
\}$, then we use ($\ref{unstable}$) to conclude that an initial
application of $s( \lambda)$ followed by iteration of
$g_0(\lambda)$ gives an infinite orbit. This shows ($\ref{3dir}$)
and completes the proof for energies in a spectral gap.

We can now complete the proof by taking $M:= M_1 \cup M_2 \cup
M_3$, where
\begin{equation*}
M_1:= \bigcup_{(a,b)} M^{(a,b)},
\end{equation*}
the union being taken over all stability intervals in
$\sigma(H_0)$,
\begin{equation*}
M_2:= \bigcup_{(\alpha,a)} M^{( \alpha, a)},
\end{equation*}
the union being taken over all maximal gaps of $H_0$, and $M_3$
defined to be the set containing all the endpoints of the
stability intervals. \hfill \qed

\setcounter{equation}{0}


\section{H\"{o}lder Continuity of the Lyapunov Exponent}\label{lyapuhoelder}

In this section, we prove that the Lyapunov exponent for the
family of random operators defined by $( \ref{contam})$, $( \ref{apot})$ is H\"{o}lder
continuous on every compact interval of $\R$ which doesn't contain energies
in the exceptional set $M$ from Theorem \ref{gamma+}. A result of this type is proven in
\cite{Carmona/Lacroix} for an analogue of ($\ref{discreteam}$) defined on
a discrete strip. We give an outline of the proof indicating what
changes are necessary in order to see that the argument from
\cite{Carmona/Lacroix} can be adapted to our setting. The main
changes are due to less explicit forms of the continuum transfer
matrices.

Fix a compact interval $I \subset \R \setminus M$, where $M$ is the
discrete set from Theorem \ref{gamma+}.

\begin{Theorem} \label{hcog} The Lyapunov exponent for $( \ref{contam})$,  $( \ref{apot})$ is
uniformly H\"{o}lder continuous on $I$, i.e. there exists a number $\alpha>0$ and a
constant $C$ for which
\begin{equation*}
| \gamma( \lambda) - \gamma( \lambda^{\prime})| \leq C | \lambda - \lambda^{\prime}|^{\alpha},
\end{equation*}
for all $\lambda, \lambda^{\prime} \in I$.
\end{Theorem}

The rest of this section will be used to prove Theorem~\ref{hcog},
where we rely on general facts from Appendix B.

As in the introduction let $g_{\lambda}(n,\omega)$ be the transfer
matrix of ($\ref{schrod}$) from $n - 1/2$ to $n + 1/2$. If $u_N( \cdot,
\lambda, \omega)$ and $u_D( \cdot, \lambda, \omega)$ are the
solutions of ($\ref{schrod}$) with $u_N(n-1/2)=u_D^{\prime}(n-1/2)=1$,
$u_N^{\prime}(n-1/2)=u_D(n-1/2)=0$, then
\begin{equation} \label{transfer}
g_{\lambda}(n, \omega) = \left( \begin{array}{cc} u_N( n+ 1/2, \lambda, \omega) & u_D( n+ 1/2, \lambda, \omega) \\
u^{\prime}_N( n+ 1/2, \lambda, \omega) & u^{\prime}_D( n+ 1/2, \lambda, \omega) \end{array}
\right).
\end{equation}
This shows that for fixed $n$ and $\lambda$, $g_{\lambda}( n,
\cdot): \Omega \rightarrow {\rm SL}(2, \R)$ is measurable, and for fixed
$\lambda$, $g_{\lambda}(n, \omega)$ are real-valued, i.i.d.\ random
matrices. Here we have used supp$(f) \subset [-1/2,1/2]$.

Let $\mu_{\lambda}$ denote the distribution of $g_{\lambda}$ in
${\rm SL}(2,\R)$. Using ($\ref{transfer}$) and the boundedness of the
distribution $\mu$ of $q_n$, we get from Lemma~\ref{growthbound} that

\begin{equation} \label{gron1}
\M g_{\lambda}(n, \omega) \M^2 \leq
\mbox{exp}(C_1+|\lambda|+C_2|q_n(\omega)|) \leq C_3,
\end{equation}
and from Lemma~\ref{diffbound}
\begin{equation} \label{gron2}
\M g_{\lambda}(n, \omega) - g_{\lambda^{\prime}}(n, \omega) \M \leq C|\lambda- \lambda^{\prime}|,
\end{equation}
where all constants are uniform in $\omega$,$n$, and $\lambda,
\lambda^{\prime} \in I$.

Take $G= {\rm SL}(2, \R)$, $B= P( \R^2)$, and the action $g \overline{v} = \overline{gv}$ as
described in Section~\ref{lyapunovsec}.
The projective distance
\begin{equation*}
\delta( \overline{x}, \overline{y}) = \frac{|x_1y_2-x_2y_1|}{\M x \M \cdot \M y \M}, \ \ \mbox{for } x,y \in \R^2
\end{equation*}
defines a metric on $P( \R^2)$. (Note that $\delta( \overline{x},
\overline{y})=| \sin( \psi)|$, where $\psi$ is the angle between
$x$ and $y$.) As in Appendix B, Example~\ref{ex}, set $\tilde{B}:=
B \times B \setminus \{ (b,b): b \in B \}$ and define a cocycle
$\sigma_1: G \times \tilde{B} \rightarrow (0, \infty)$ by
\begin{equation} \label{sigma}
\sigma_1 \left( g, ( \overline{x}, \overline{y}) \right) :=
\frac{\delta(g \overline{x}, g \overline{y})}{\delta( \overline{x}, \overline{y})}.
\end{equation}
A short calculation shows that
\begin{equation*}
\sigma_1 \left( g, ( \overline{x}, \overline{y}) \right) =
\frac{ \M x \M \cdot \M y \M }{ \M g x \M \cdot \M g y \M },
\end{equation*}
and hence
\begin{equation*}
\overline{\sigma}_1(g):= \sup_{b \in \tilde{B}} \sigma_1(g, b) \leq \M g \M^2,
\end{equation*}
thus ($\ref{gron1}$) implies that
$\overline{\sigma}_1(g) \leq e^{C+| \lambda|}$ for $\mu_{\lambda}$-a.e.\ $g$. In particular,
this shows that for every $t \in \R$,
\begin{equation*}
\sup_{\lambda \in I} \int \overline{ \sigma}_1(g)^t d \mu_{\lambda}(g) < \infty.
\end{equation*}
This proves that for ($\ref{sigma}$), assumption (i) of
Lemma~\ref{cocycle} is satisfied uniformly with respect to
$\lambda \in I$. Proving that assumption (ii) of
Lemma~\ref{cocycle} also holds for ($\ref{sigma}$), again
uniformly with respect to $\lambda \in I$, requires more work.

To this end, for any $\overline{v} \in P( \R^2)$, define
\begin{equation*}
\Phi_{\lambda}( \overline{v}):= \E \left\{ \log \frac{ \M g_{\lambda} v \M} { \M v \M } \right\}.
\end{equation*}
The next lemma corresponds to Lemma V.4.7 of \cite{Carmona/Lacroix}.
\begin{Lemma} {\rm (i)} The mapping $\Phi_{\lambda}:P( \R^2) \rightarrow \R$ is
continuous.

\noindent {\rm (ii)} The mapping $\Phi: I \times P(\R^2) \rightarrow \R$ with
$\Phi( \lambda, \overline{v})=\Phi_{\lambda}( \overline{v})$ satisfies
\begin{equation} \label{cophi}
\sup_{ \overline{v} \in P(\R^2)} | \Phi_{\lambda}( \overline{v}) - \Phi_{\lambda^{\prime}}( \overline{v})|
\leq C| \lambda - \lambda^{\prime}|,
\end{equation}
for some constant $C=C(I)$. In particular, $\Phi$ is continuous.
\end{Lemma}
\begin{proof}
(i) By ($\ref{gron1}$), $\log \M g_{\lambda}v \M / \M v \M \leq \log \M g_{\lambda} \M \leq C$
uniformly in $\omega$ and $v$.  Thus by dominated convergence,
$\lim_{n \rightarrow \infty} \Phi_{\lambda}( \overline{v}_n) = \Phi_{\lambda}( \overline{v})$,
if $\overline{v}_n \rightarrow \overline{v}$.

(ii) We have that
\begin{equation*}
| \Phi_{\lambda}( \overline{v}) - \Phi_{\lambda^{\prime}}( \overline{v})| =
\left| \E \left\{ \log \frac{ \M g_{\lambda} v \M } { \M g_{\lambda^{\prime}} v \M } \right\} \right|
\leq \E \left\{ \log \M g_{\lambda} g_{\lambda^{\prime}}^{-1} \M \right\}
\end{equation*}
\begin{equation*}
\leq \E \left\{ \log \left( \M g_{\lambda} - g_{\lambda^{\prime}} \M \cdot \M g_{\lambda^{\prime}}^{-1} \M +1 \right) \right\}
\leq C| \lambda - \lambda^{\prime}|,
\end{equation*}
where the last inequality follows as $\M g_{\lambda} - g_{\lambda^{\prime}} \M \cdot \M g_{\lambda^{\prime}}^{-1} \M
\leq C| \lambda - \lambda^{\prime}|$ for all $\lambda, \lambda^{\prime} \in I$ by ($\ref{gron1}$)
and ($\ref{gron2}$).
\end{proof}

As we know that for each fixed $\lambda \in I$, $G( \lambda)$ is non-compact and strongly irreducible,
then combining Proposition IV.4.11 and Theorem IV.4.14 of \cite{Carmona/Lacroix} we have that
there exists a unique $\mu_{\lambda}$-invariant probability measure $\nu_{\lambda}$ on
$P( \R^2)$ for which
\begin{equation} \label{geqn}
\gamma( \lambda) = \int \log \frac{ \M gv \M}{ \M v \M} d \mu_{\lambda}(g) d \nu_{\lambda}( \overline{v}).
\end{equation}

\begin{Corollary}\label{coro3.3}

\noindent {\rm (i)} $\gamma$ is continuous on $I$, and

\noindent {\rm (ii)}
\begin{equation*}
\lim_{n \rightarrow \infty} \frac{1}{n} \E \left\{ \log \frac{ \M
U_{\lambda}(n) v \M }{ \M v \M } \right\} = \gamma( \lambda),
\end{equation*}
uniformly with respect to $\lambda \in I$ and $\overline{v} \in P( \R^2)$.
\end{Corollary}
\begin{proof}
\noindent (i) We first show that $\nu_{\lambda^{\prime}}
\rightarrow \nu_{\lambda}$ weakly as $\lambda^{\prime} \rightarrow
\lambda$. Note that in general if $\mu_{\lambda_n} \rightarrow
\mu_{\lambda_0}$ weakly and $\nu_{\lambda_n} \rightarrow
\nu_{\lambda_0}$ weakly, then $\mu_{\lambda_n}*\nu_{\lambda_n}
\rightarrow \mu_{\lambda_0}*\nu_{\lambda_0}$ weakly (for the definition of the convolution, see
Appendix B). By
$(\ref{gron2})$ we have that $\mu_{\lambda^{\prime}} \rightarrow
\mu_{\lambda}$ weakly and an application of Arzela-Ascoli implies
that every subsequence of $\{ \nu_{\lambda^{\prime}} \}$ contains
a weakly convergent subsequence. Using this, the convergence of
the convolutions, and the uniqueness of the invariant measures, we
see that $\nu_{\lambda^{\prime}} \rightarrow \nu_{\lambda}$. Now,
continuity follows from ($\ref{geqn}$), i.e. noting that $\gamma(
\lambda) = \nu_{\lambda}( \Phi_{\lambda} )$, and the estimate
($\ref{cophi}$), see Corollary V.4.8 of \cite{Carmona/Lacroix}.

\noindent (ii) This follows as in the proof of ($\ref{cophi}$),
with ($\ref{gron1}$) and ($\ref{gron2}$) replaced by corresponding
estimates for $U_{\lambda}(n,1)$, which again follow from
Lemmas~\ref{growthbound} and \ref{diffbound}, see
Proposition~V.4.9 of \cite{Carmona/Lacroix}.
\end{proof}

\begin{Lemma} \label{cocycle1} There exists an integer $N$ for which
\begin{equation*}
\sup_{\lambda \in I, b \in \tilde{B}} \int \log \left[ \sigma_1(g,b) \right] d \mu_{\lambda}^N(g) <0.
\end{equation*}
\end{Lemma}
\begin{proof} One calculates that
\begin{equation*}
\frac{1}{n} \int \log \frac{ \delta( g \overline{x}, g \overline{y})}{ \delta( \overline{x}, \overline{y})} d \mu_{\lambda}^n(g)
= - \frac{1}{n} \int \log \frac{ \M g x \M}{ \M x \M }d \mu_{\lambda}^n(g)
- \frac{1}{n} \int \log \frac{ \M g y \M}{ \M y \M }d \mu_{\lambda}^n(g)
\end{equation*}
\begin{equation*}
= - \frac{1}{n} \E \left\{ \log \frac{ \M U_{\lambda}(n) x \M}{ \M
x \M } \right\} - \frac{1}{n} \E \left\{ \log \frac{ \M
U_{\lambda}(n) y \M}{ \M y \M } \right\}.
\end{equation*}
The result follows by letting $n \rightarrow \infty$ and noting both (i) and (ii) of the above Corollary.
\end{proof}

We have thus shown that Lemma~\ref{cocycle} is applicable to the
cocycle $\sigma_1$ defined in (\ref{sigma}). Following the lines
of \cite{Carmona/Lacroix}, this leads to H\"older continuity
properties with respect to $\lambda$ of the invariant measures
$\nu_{\lambda}$ and the operators $T_{\lambda}$ defined by
$(T_{\lambda}f)(\overline{x}) = \int
f(g\overline{x})\,d\mu_{\lambda}(g)$ for $f$ in the H\"older
spaces $\Cal{L}_{\alpha}(P(\R^2))$, see Appendix B.

\begin{Lemma}\label{L3.7}
There exists an $\alpha_0>0$ such that for $0< \alpha \leq \alpha_0$, there exists
$\rho_{\alpha}<1$ and $C_{\alpha} < \infty$ such that:
\begin{equation*}
\sup_{ \lambda \in I, \overline{x} \neq \overline{y} \in P( \R^2)}
\E \left\{ \frac{ \delta(U_{\lambda}(n) \overline{x},
U_{\lambda}(n) \overline{y})^{\alpha}}{ \delta( \overline{x},
\overline{y})^{\alpha}} \right\} \leq C_{\alpha} \rho_{\alpha}^n,
\end{equation*}
and
\begin{equation*}
\sup_{ \lambda \in I} \M T^n_{\lambda}f - \nu_{\lambda}(f) \M \leq C_{\alpha} \M f \M_{\alpha} \rho_{\alpha}^n,
\end{equation*}
for $n=1,2, . . .$
\end{Lemma}
\begin{proof} Lemma \ref{cocycle1} confirms that both assumptions of
Lemma \ref{cocycle} hold uniformly with respect to $\lambda \in
I$. That this implies the above estimates, again uniformly with
respect to $\lambda \in I$, follows from observing that the proofs
of Propositions IV.3.5 and IV.3.15 yield uniform results under
uniform assumptions.
\end{proof}
\begin{Lemma}
There exists an $\alpha_0>0$ such that for $0< \alpha \leq \alpha_0$, there exists
a $C_{\alpha} < \infty$ such that:

\noindent {\rm (i)}
\begin{equation*}
\M T_{\lambda}f - T_{\lambda^{\prime}}f \M_{\alpha/2} \leq C_{\alpha} \M f \M_{\alpha}
| \lambda- \lambda^{\prime}|^{\alpha/2},
\end{equation*}

and

\noindent {\rm (ii)}
\begin{equation*}
| \nu_{\lambda}(f) - \nu_{\lambda^{\prime}}(f) | \leq C_{\alpha} \M f \M_{\alpha}
| \lambda- \lambda^{\prime}|^{\alpha/2},
\end{equation*}
for  all $\lambda, \lambda^{\prime} \in I$ and $f \in \Cal{L}_{\alpha}$.
\end{Lemma}
\begin{proof}
(i) follows as in V.4.13 of \cite{Carmona/Lacroix} using that
\begin{equation*}
\delta(g_{\lambda} \overline{x}, g_{\lambda^{\prime}} \overline{x}) =
\frac{ \left| \left( (g_{\lambda} -g_{\lambda^{\prime}})x \right)_1 (g_{\lambda^{\prime}}x)_2 +
(g_{\lambda^{\prime}}x)_1 \left( (g_{\lambda^{\prime}} -g_{\lambda})x \right)_2 \right|}
{ \M g_{\lambda}x \M \M g_{\lambda^{\prime}}x \M}
\end{equation*}
\begin{equation*}
\leq 2 \cdot \frac{ \M (g_{\lambda} -g_{\lambda^{\prime}}) x \M}{ \M g_{\lambda}x \M}
\leq 2 \cdot \M g_{\lambda} -g_{\lambda^{\prime}} \M \cdot \M g_{\lambda}^{-1} \M
\leq C | \lambda - \lambda^{\prime}|.
\end{equation*}

\noindent (ii) follows as in Proposition V.4.14 of
\cite{Carmona/Lacroix} using Lemma \ref{cocycle} (c) as done
therein.
\end{proof}
Compiling these results, Theorem $\ref{hcog}$ follows as in Theorem V.4.15
of \cite{Carmona/Lacroix}.

\setcounter{equation}{0}


\section{H\"{o}lder Continuity of the Integrated Density of States}

In this section we prove an analogue of Theorem $\ref{hcog}$ for the integrated density
of states. We begin with its definition. Let, for $L \in \N$ and
$\Lambda_L(0)= [-L/2,L/2]$,
\begin{equation}
H_{\Lambda_L(0)}( \omega):=H_{\omega} \ \ \mbox{on} \ \ L^2(\Lambda_L(0))
\end{equation}
with $H_{\omega}$ as in $( \ref{contam})$, $(\ref{apot})$ and
Dirichlet boundary conditions set at $\pm L/2$. Fix $\lambda \in
\R$ and let $N_{L, \omega}(\lambda)$ denote the number of
eigenvalues of $H_{\Lambda_L(0)}( \omega)$ less than or equal to
$\lambda$. By Kingman's super-additive ergodic theorem, we have
that the following limit
\begin{equation} \label{ids}
N( \lambda):= \lim_{L \rightarrow \infty} \frac{ N_{L, \omega}(\lambda)}{L}
\end{equation}
exists for almost every $\omega$ and in expectation. In particular,
\begin{equation} \label{eids}
N(\lambda) = \sup_{L \geq1} \frac{1}{L} \E \left( N_{L, \omega}( \lambda) \right).
\end{equation}
The limit $N(\lambda)$ in $(\ref{ids})$ does not depend on the
particular choice of boundary conditions one sets on
$\Lambda_L(0)$, however, the monotonicity of the limit in
expectation $(\ref{eids})$ may be changed, e.g. with Neumann
boundary conditions one gets an ``$\inf$'' rather than a ``$\sup$.''
$N(\lambda)$ is called the integrated density of states.

As in Section \ref{lyapuhoelder}, fix a compact interval $I \subset \R
\setminus M$, $M$ as in Theorem \ref{gamma+}.

\begin{Theorem}\label{hcoids}
The integrated density of states for \eqref{contam}, \eqref{apot} is uniformly H\"{o}lder continuous on $I$.
\end{Theorem}

The proof of this theorem follows easily from Theorem $\ref{hcog}$ once we have a means
of relating $N$ and $\gamma$. For our model, one has

\begin{Proposition}{\rm (The Thouless formula)} \label{thouless}
Let $N$ and $\gamma$ respectively be the integrated density of
states and Lyapunov exponent for \eqref{contam}, \eqref{apot}.
Then there exists $\alpha \in \R$ such that, for every $\lambda
\in \R$,
\begin{equation} \label{thouform}
\gamma( \lambda) = - \alpha + \int_{\R} \log \left| \frac{ \lambda - t}{t-i} \right| dN(t).
\end{equation}
\end{Proposition}

{\bf Remarks:} (i) The above version of the Thouless formula
arises in Kotani's work, e.g. \cite{Kotani}. It differs from the
discrete version, see Proposition VI.4.3 of
\cite{Carmona/Lacroix}, essentially by the normalization term
$t-i$ in (\ref{thouform}), which compensates for the non-compact
support of $dN$. Alternatively, one can work with the normalized
Lyapunov exponent $\gamma-\gamma_0$ and integrated density of
states $N-N_0$, where $\gamma_0$ and $N_0$ are the corresponding
`free' quantities, see \cite{Avron/Simon}. We decided to work with
(\ref{thouform}) since it was shown in Appendix 2 of
\cite{Kirsch/Kotani/Simon} that Kotani theory applies to singular
potentials. While \cite{Kirsch/Kotani/Simon} works with
$L^2$-assumptions on the potential, it is pointed out that all
what is needed to make Kotani theory work are two basic estimates
for the $m$-function $m(z, \omega)$ on $(0, \infty)$ of an ergodic
random operator: If $K$ is a compact subset of $\C^+$, the
complex, upper-half plane, then there are constants $C(K)
< \infty$ and $\delta(K)> 0$ such that
\begin{equation} \label{m1}
|m(z, \omega)| \leq C(K),
\end{equation}
and
\begin{equation} \label{m2}
| {\rm Im}[m(z, \omega)]| \geq \delta(K),
\end{equation}
uniformly for $z \in K$ and $\omega \in \Omega$. In our model
(\ref{contam}), (\ref{apot}) we have only used
$L^1_{{\rm loc}}$-assumptions on the potentials. That (\ref{m1}) is
satisfied in this case follows from results on $m$-function
asymptotics in \cite{Gesztesy/Simon}. To get (\ref{m2}) we use
that
\begin{equation} \label{imform}
{\rm Im}[m(z)] = {\rm Im}[z] \cdot \int_0^{\infty}|u|^2dx,
\end{equation}
where $u$ is the solution of $-u^{\prime \prime}
+(V_{{\rm per}}+V_{\omega})u= zu$ with $u(0)=1$ and $u'(0)=m(z)$, e.g.
\cite{Codd/Lev}. Combining Lemma A.1 and A.3, we get
\begin{equation*}
\int_0^2|u|^2dx \geq C_1 \left( |u(1)|^2+|u'(1)|^2 \right) \geq
C_1C_2 \left( 1 + |m(z)|^2 \right).
\end{equation*}
Thus (\ref{imform}) yields (\ref{m2}).

(ii) To apply Kotani's results from \cite{Kotani}, one must have
an $\R$-ergodic system, i.e. there must exist a group $\{
\theta_t: t \in \R \}$ of measure preserving transformations for
which the dynamical system $( \Omega, \Cal{F}, \theta_t, \mu)$ is
ergodic and the random potential $v_{\omega}$ satisfies
$v_{\theta_t \omega}(x)=v_{\omega}(x+t)$. Kirsch's result in
\cite{Kirsch} shows how a $\Z$-ergodic system, i.e. one for which
the transformations $\theta_t$ are parametrized by $t \in \Z$, can
be associated with an $\R$-ergodic system, embedded in a larger
probability space. The associated system is constructed in such a
way that the corresponding integrated density of states and
Lyapunov exponent for both systems are equal. Our model is
$\Z$-ergodic when equipped with translations. Thus we must apply
Kotani's results to the corresponding $\R$-ergodic system after
using Kirsch's suspension procedure.

(iii) In Kotani's work \cite{Kotani}, $N$ and $\gamma$ arise as
the real and imaginary parts of the non-tangential limit of a
specific Herglotz function (the $w$-function). To see that the
integrated density of states and the Lyapunov exponent actually
coincide with these non-tangential limits, see Proposition V.12
and Proposition VI.1 of \cite{CLN}.

\vspace{3mm}

Based on (\ref{thouform}), the proof of Theorem \ref{hcoids} is
very similar to the proof of the corresponding result in the
discrete case, see Proposition VI.3.9 of \cite{Carmona/Lacroix}.
Some small changes arise from the use of the slightly different
Thouless formula (\ref{thouform}) and the necessity to exclude the
set $M$. We use some basic facts about Hilbert transforms, which
for a square integrable function $\psi$ is defined by:
\begin{equation*}
(T \psi)(x) = \frac{1}{\pi} \lim_{\epsilon \rightarrow 0^+} \int_{|x-t|> \epsilon} \frac{ \psi(t)}{x-t}dt.
\end{equation*}
One has two basic results, as stated in \cite{Carmona/Lacroix},
\newline (i) $T \psi$ is a square integrable function and $T^2 \psi = - \psi$ almost everywhere
with respect to Lebesgue measure.
\newline (ii) If $\psi$ is H\"{o}lder continuous on some interval $[-a,a]$, then $T \psi$ is
H\"{o}lder continuous on $[- \frac{a}{2}, \frac{a}{2}]$.

Note that if $\psi_a(t):= \psi(t-a)$, then $T \psi_a = (T
\psi)_a$. Thus if $\psi$ is H\"{o}lder continuous on some interval
$[x_0-a,x_0+a]$, then $T \psi$ is H\"{o}lder continuous on $[x_0-
\frac{a}{2}, x_0+ \frac{a}{2}]$.

\medskip
{\noindent}{\it Proof of Theorem \ref{hcoids}.} First, observe that
$dN$-integrability of $\log|( \lambda-t)/(t-i)|$ yields
\begin{equation*}
\lim_{\epsilon \rightarrow 0^+} \int_{\lambda - \epsilon}^{\lambda
+ \epsilon} \left| \log | \frac{ \lambda-t}{t-i}| \right|dN(t) =0,
\end{equation*}
which easily implies that
\begin{equation} \label{lhc}
\lim_{\epsilon \rightarrow 0^+} | \log( \epsilon)| \left( N(
\lambda + \epsilon)-N(\lambda - \epsilon) \right) =0,
\end{equation}
i.e. log-H\"older continuity and, in particular, continuity of
$N$.

Let $\lambda_0 \in I$ and pick $a>0$ such that $[ \lambda_0 - 4a,
\lambda_0 + 4a] \subset \R \setminus M$ and thus $\gamma$ is
H\"older continuous in $[ \lambda_0 - 4a, \lambda_0 + 4a]$. Take
$\psi(t):= N(t) \chi_{ \{t: |t- \lambda_0| \leq 4a \}}$, and note
that $(T^2 \psi)(t) = - N(t)$ for almost every $t$ with $|t -
\lambda_0| \leq 4a$ by $(i)$. For $| \lambda - \lambda_0|<4a$
calculate,
\begin{align*}
\gamma( \lambda) + \alpha - \int_{|t - \lambda_0|>4a} & \log
\left| \frac{ \lambda - t}{t-i} \right| dN(t) = \int_{ \lambda_0
-4a}^{ \lambda_0 + 4a} \log \left| \frac{ \lambda - t}{t-i}
\right| dN(t)
\\ = & \lim_{ \epsilon \rightarrow 0^+} \left( \int_{ \lambda +
\epsilon}^{ \lambda_0 + 4a} \log | \lambda - t| dN(t) + \int^{
\lambda - \epsilon}_{ \lambda_0 - 4a} \log | \lambda - t | dN(t)
\right)\\ & - \frac{1}{2} \int_{ \lambda_0 - 4a}^{ \lambda_0 + 4a}
\log (t^2+1)dN(t).
\end{align*}

Integrating the first two integrals above by parts and rearranging yields

\begin{equation} \label{hton}
\pi(T \psi)( \lambda) = \gamma( \lambda) + \alpha - \int_{|t-
\lambda_0|
> 4a} \log \left| \frac{ \lambda - t}{t-i} \right| dN(t) +
\frac{1}{2} \int_{ \lambda_0 - 4a}^{ \lambda_0 + 4a} \log
(t^2+1)dN(t)
\end{equation}
\begin{equation*}
 - \log( \lambda_0 - \lambda +4a)N( \lambda_0 + 4a) + \log( \lambda - \lambda_0 +4a)N( \lambda_0 - 4a),
\end{equation*}
where (\ref{lhc}) has been used. From observing (\ref{hton}), we
see that $T \psi$ is H\"{o}lder continuous in $[ \lambda_0 -2a,
\lambda_0 +2a]$, and thus $T^2 \psi$ is H\"{o}lder continuous in
$[ \lambda_0 -a, \lambda_0 +a]$ by (ii). Since $N$ is continuous
we get $(T^2 \psi)(t) = -N(t)$ for all $t \in [ \lambda_0 -a,
\lambda_0 +a]$. Thus $N$ is H\"older continuous in $[ \lambda_0
-a, \lambda_0 +a]$. A compactness argument yields uniform
H\"{o}lder continuity over all of $I$. \hfill \qed

\setcounter{equation}{0}


\section{The Wegner Estimate}

In this section we prove a Wegner estimate which constitutes one
of the two ingredients that will enable us to start the multiscale
induction. Given H\"older continuity of the integrated density of
states, as proven in the preceding section, our proof of the
Wegner estimate can be carried out in a way analogous to
\cite{CKM}. Since a few modifications of the arguments in
\cite{CKM} are required, we present sufficiently many details for
the reader's convenience.

Fix throughout this section a compact interval $I = [a,b] \subset
\R \setminus M,$ where $M$ is the discrete set found in Theorem
\ref{gamma+}. Our goal will be to prove estimates uniformly in
$I$. It will therefore be convenient to have these two properties
in a ball of fixed radius around each point in $I$. By our above
results we know that there is some $\xi
> 0$ such that for every $\lambda \in I_\xi = [a-\xi , b+\xi]$,
$G(\lambda)$ is non-compact and strongly irreducible. In other
words, for every $\lambda \in I$, we have that $G(\lambda')$ is
non-compact and strongly irreducible for every $\lambda' \in
[\lambda - \xi , \lambda + \xi]$.

Let, for odd $L \in \N$ and $\Lambda = \Lambda_L(0) = [-L/2,L/2]$,
$H_\Lambda(\omega)$ be the restriction of $H_\omega$ to $\Lambda$
with Dirichlet boundary conditions at $-L/2$ and $L/2$. We will
prove the following theorem:

\begin{Theorem}[Wegner estimate]\label{wegner}
For every $\beta \in (0,1)$ and every $\sigma > 0$, there exist
$L_0 \in \N$ and $\alpha > 0$ such that

\begin{equation}\label{weges}
\p \{ \dist (\lambda,\sigma( H_\Lambda(\omega)) \le e^{-\sigma
L^\beta} \} \le e^{- \alpha L^\beta}
\end{equation}
for all $\lambda \in I$ and $L \ge L_0$.
\end{Theorem}

To prove Theorem \ref{wegner}, we will need two lemmas,
Lemma~\ref{weglem1} and Lemma~\ref{weglem2} below.

\begin{Lemma}\label{weglem1}
There exist $\alpha_1 > 0, \delta > 0, n_0 \in \N$ such that for
all $\lambda \in I, n \ge n_0$, and $x$ normalized, we have $$ \E
\{ \| U_\lambda(n)x \|^{-\delta} \} \le e^{- \alpha_1 n}. $$
\end{Lemma}

\begin{proof}
This lemma can be proved in exactly the same way Lemma~5.1 is
proved in \cite{CKM}. For the reader's convenience, we sketch the
argument briefly. By our choice of the interval $I$ and the
results from Sections~\ref{lyapunovsec} and \ref{lyapuhoelder}, in
particular Corollary~\ref{coro3.3}~(i), we have $\gamma := \inf \{
\gamma(\lambda) : \lambda \in I\} > 0$. Using the inequality $e^y
\le 1 + y + y^2 e^{|y|}$ and H\"older's inequality, one shows as
in \cite{CKM} that for every $\lambda \in I$ and every $\delta >
0$, we have

\begin{align*}
\E \{ \| U_{\lambda} (n,\omega ) x \|^{-\delta} \} = & \, \E \{ \|
g_{\lambda} (n,\omega ) \cdot \ldots \cdot g_{\lambda} (1,\omega )
x \|^{-\delta} \} \\ \le & \, 1 - \delta \E \{ \log \| g_\lambda
(n,\omega) \cdot \ldots \cdot g_\lambda (1,\omega) x \| \}\\ & \,
+ \delta^2 n \left[ \E \{ ( \log \| g_\lambda (1,\omega) \|)^4 \}
\right]^{1/2} \left[ \E \{ \| g_\lambda (1,\omega) \|^{2 \delta}
\} \right]^{n/2}\\ \le & \, 1 - \delta \E \{ \log \| g_\lambda
(n,\omega) \cdot \ldots \cdot g_\lambda (1,\omega) x \| \} +
\delta^2 n^2 C_1 C_2^n
\end{align*}
for some finite constants $C_1 = C_1(I)$, $C_2 = C_2(I)$. Hence,
by Corollary~\ref{coro3.3}~(ii), we have for some $n_0 = n_0(I)$,
uniformly in $\lambda \in I$ and $x$ in the unit sphere,

\begin{equation}\label{anchor}
\E \{ \| U_{\lambda} (n_0,\omega ) x \|^{-\delta} \} \le 1 -
\tfrac{1}{2} n_0 \delta \gamma + \delta^2 n_0^2 C_1 C_2^{n_0} \le
1 - \varepsilon
\end{equation}
for some $\varepsilon > 0$, provided $\delta$ is small enough.
Iterating \eqref{anchor} as in \cite{CKM} yields $$ \E \{ \|
U_{\lambda} (n,\omega ) x \|^{-\delta} \} \le C ( 1 - \varepsilon
)^{\lfloor n/n_0 \rfloor} \le e^{-\alpha_1 n} $$ for all $n \ge
n_1$ and $\lambda \in I$, for some $\alpha_1 = \alpha_1(I) > 0$
and $n_1 = n_1(I)$, where $\lfloor n/n_0 \rfloor$ is the largest
integer less or equal $n/n_0$.
\end{proof}

\begin{Lemma}\label{weglem2}
There exist $\rho > 0$ and $C < \infty$ such that for every
$\lambda \in I$ and every  $\varepsilon > 0$, we have for $L \ge
L_0$,

\begin{align}\label{smallprob}
\p \{ & \mbox{There exist } \lambda' \in (\lambda -
\varepsilon,\lambda + \varepsilon) \mbox{ and } \phi \in
D(H_\Lambda), \, \|\phi\|=1, \mbox{ such that}\\ \nonumber &
(H_\Lambda (\omega) - \lambda')\phi = 0, \, |\phi'(-L/2)|^2 +
|\phi'(L/2)|^2 \le \varepsilon^2 \} \le C L \varepsilon^\rho.
\end{align}
\end{Lemma}

\begin{proof}
We will follow the same strategy as Carmona et al.\ in their proof
of \cite[Lemma~5.2]{CKM}, that is, we will use H\"older continuity
of the integrated density of states to derive the estimate
\eqref{smallprob}. The only difficulty that arises is that the
cutoff of eigenfunctions as performed by Carmona et al.\ in the
discrete case may produce elements outside the domain of the local
Hamiltonian. We therefore use a smooth cutoff procedure and show
that the argument still goes through.

Note first that it suffices to prove \eqref{smallprob} for small
$\varepsilon > 0$. It follows from Theorem~\ref{hcoids} that there
are constants $\rho = \rho(I_\xi) > 0$ and $C_1 = C_1(I_\xi) <
\infty$ such that for every $\lambda, \lambda' \in I_\xi$,

\begin{equation}\label{smoothids}
| N(\lambda) - N(\lambda') | \le C_1 | \lambda - \lambda' |^\rho.
\end{equation}

Now fix $\lambda \in I$, $\varepsilon > 0$, and $L$. Let, for $k
\in \Z$, $\Lambda_k$ be the interval $[kL - L/2, kL + L/2]$ and
denote by $H_{\Lambda_k}(\omega)$ the operator $H(\omega)$
restricted to $\Lambda_k$ with Dirichlet boundary conditions.

Let $A_k$ be the event $A_k = A_k(\lambda,\varepsilon,L) = \{
\omega \in \Omega : H_{\Lambda_k}(\omega)$ has an eigenvalue
$\lambda_k \in (\lambda - \varepsilon, \lambda + \varepsilon)$
such that the corresponding normalized eigenfunction $\phi_k$
satisfies $| \phi_k'(kL - L/2) |^2 + | \phi_k'(kL + L/2) |^2 \le
\varepsilon^2 \}$. Let $p = p(\lambda, \varepsilon, L) = \p \{ A_k
\}$. Clearly, $p$ is independent of $k$ and equals the left hand
side of \eqref{smallprob}.

Fix some $n \in \N$ and let $H_n(\omega)$ be the operator
$H(\omega)$ restricted to $\Lambda(n) := \bigcup_{k=-n}^n
\Lambda_k = [-nL - L/2, nL + L/2]$ with Dirichlet boundary
conditions at $-nL -L/2$ and $nL + L/2$. Let $k_1, \ldots ,k_j \in
\{-n, \ldots , n\}$ be distinct and such that the event $A_{k_l}$
occurs. For each such $l$, we shall construct a normalized
function $\tilde{\phi}_l$ in the domain of $H_n(\omega)$ which is
supported on $\Lambda_{k_l}$ (hence $\{ \tilde{\phi}_1, \ldots ,
\tilde{\phi}_j \}$ form an orthonormal set) such that for every
$l$,

\begin{equation}\label{smallnorm}
\| (H_n(\omega) - \lambda) \tilde{\phi}_l \| \le C_2 \varepsilon,
\end{equation}
where $C$ is a constant which only depends on the single site
potential $f$ and the single site distribution $\mu$. By
disjointness of supports, we also have

\begin{equation}\label{ortho}
\langle \tilde{\phi}_l , H_n(\omega) \tilde{\phi}_{l'} \rangle = 0
= \langle H_n(\omega) \tilde{\phi}_l , H_n(\omega)
\tilde{\phi}_{l'} \rangle \mbox { for } l \not= l'.
\end{equation}

By \cite[Lemma~A.3.2]{st}, \eqref{smallnorm} and \eqref{ortho}
imply that the number of eigenvalues of $H_n(\omega)$ (counted
with multiplicity) in $[\lambda - C_2 \varepsilon, \lambda + C_2
\varepsilon]$ is bounded from below by $j$. In other words, $$ \#
\{ l \in \{-n, \ldots, n\} : A_l \mbox{ occurs} \} \le \# \{
\mbox{eigenvalues of } H_n(\omega) \mbox{ in } [\lambda - C_2
\varepsilon , \lambda + C_2 \varepsilon] \}. $$ Thus, if
$\varepsilon$ is small enough (namely, such that $[\lambda - C_2
\varepsilon , \lambda + C_2 \varepsilon] \subseteq I_\xi$ or,
equivalently, $\varepsilon \le \xi/C_2$), we have

\begin{align*}
p = & \, \lim_{n \rightarrow \infty} \frac{1}{2n+1} \# \{ l \in
\{-n, \ldots, n\} : A_l \mbox{ occurs} \}\\ \le & \, \lim_{n
\rightarrow \infty} \frac{1}{2n+1} \# \{ \mbox{eigenvalues of }
H_n(\omega) \mbox{ in } [\lambda - C_2 \varepsilon , \lambda + C_2
\varepsilon] \}\\ = & \, L \lim_{n \rightarrow \infty}
\frac{1}{(2n+1)L} \# \{ \mbox{eigenvalues of } H_n(\omega) \mbox{
in } [\lambda - C_2 \varepsilon , \lambda + C_2 \varepsilon] \}\\
= & \, L (N(\lambda + C_2 \varepsilon) - N(\lambda - C_2
\varepsilon))\\ \le & \, L C_1 (2C_2 \varepsilon)^\rho\\ =: & \, C
L \varepsilon^\rho,
\end{align*}
where the intermediate steps hold true for almost every $\omega$.

It remains to construct $\tilde{\phi}_l$ with the desired
properties. Fix some $l$ and consider the function $\phi_l$ which
is defined on $\Lambda_{k_l}$ and vanishes at the boundary points.
In principle we would like to extend $\phi_l$ by zero on
$\Lambda(n) \setminus \Lambda_{k_l}$. However, this function will
in general not belong to the domain of $H_n$, so we could not even
evaluate $(H_n - \lambda)$ applied to this function. Instead we
will use a smooth extension of $\phi_l$ to $\Lambda(n)$.

Fix once and for all a smooth function $\chi$ which obeys $0 \le
\chi \le 1$, $\chi(x) = 0$ for $x \le 0$, and $\chi(x) = 1$ for $x
\ge 1$. Let $x_l^L = k_l L - L/2$ and $x_l^R = k_l L + L/2$ and
define $\hat{\phi}_l$ for $x \in \Lambda(n)$ by $$ \hat{\phi}_l
(x) = \left\{ \begin{array}{cll} 0 & \mbox{ if } & x \not\in
\Lambda_{k_l} \\ \chi(x - x_l^L) \phi_l(x) & \mbox{ if } &  x_l^L
\le x \le x_l^L + 1  \\ \phi_l(x) & \mbox{ if } & x_l^L + 1 \le x
\le x_l^R -1 \\ \chi( - x + x_l^R) \phi_l(x) & \mbox{ if } & x_l^R
- 1 \le x \le x_l^R \end{array} \right. $$ Then $\hat{\phi}_l$
clearly belongs to the domain of $H_n$ and it has norm bounded by
$\| \phi_l \| = 1$. We want to estimate $\| (H_n - \lambda)
\hat{\phi}_l \|$. Now $\phi_l$ is an eigenfunction corresponding
to the eigenvalue $\lambda_{k_l} \in (\lambda - \varepsilon ,
\lambda + \varepsilon)$, so we write $\| (H_n - \lambda)
\hat{\phi}_l \| \le \| (H_n - \lambda_{k_l}) \hat{\phi}_l \| +
\varepsilon$. To estimate $\| (H_n - \lambda_{k_l}) \hat{\phi}_l
\|$, we consider $$ (H - \lambda_{k_l}) \hat{\phi}_l (x) = -
\hat{\phi}_l''(x) + V_\omega(x) \hat{\phi}_l (x) - \lambda_{k_l}
\hat{\phi}_l (x) $$ for $x \in \Lambda(n)$. Hence, we have

\begin{align*}
\| (H_n - \lambda_{k_l}) \hat{\phi}_l \|^2 = & \,
\int_{x_l^L}^{x_l^L + 1} | \chi''(x - x_l^L) \phi_l(x) + 2 \chi'(x
- x_l^L) \phi_l'(x) |^2 dx\\ & \, + \int_{x_l^R - 1}^{x_l^R} |
\chi''(-x + x_l^R) \phi_l(x) + 2 \chi'(-x + x_l^R) \phi_l'(x) |^2
dx\\ \le & \, \left\| \left( \begin{array}{c} \phi_l \\ \phi_l'
\end{array} \right) \right\|_{ L^\infty(x_l^L , x_l^L + 1)}^2
\cdot \int_{x_l^L}^{x_l^L + 1} \left\| \left( \begin{array}{c}
\chi''(x - x_l^L) \\ 2 \chi'(x - x_l^L) \end{array} \right)
\right\|^2 dx\\ & \, + \left\| \left( \begin{array}{c} \phi_l \\
\phi_l' \end{array} \right) \right\|_{ L^\infty(x_l^R - 1 ,
x_l^R)}^2 \cdot \int_{x_l^R - 1}^{x_l^R} \left\| \left(
\begin{array}{c} \chi''(-x + x_l^R) \\ 2 \chi'(-x + x_l^R)
\end{array} \right) \right\|^2 dx\\ \le & \, C_3 \left[ \left\|
\left( \begin{array}{c} \phi_l(x_l^L) \\ \phi_l'(x_l^L)
\end{array} \right) \right\|^2 + \left\| \left( \begin{array}{c}
\phi_l(x_l^R) \\ \phi_l'(x_l^R) \end{array} \right) \right\|^2
\right]\\ = & \, C_3 \left[ |\phi_l'(x_l^L)|^2 +
|\phi_l'(x_l^R)|^2 \right]\\ \le & \, C_3 \varepsilon^2,
\end{align*}
where the constant $C_3$ depends on the single site potential, the
single site distribution, and the function $\chi$. Here we have
used Lemma~\ref{growthbound} in step~3. Let us define
$\tilde{\phi}_l := \hat{\phi}_l / \| \hat{\phi}_l \|$. By
construction and Lemma~\ref{growthbound}, we have, for
$\varepsilon$ sufficiently small, $\| \hat{\phi}_l \| \ge 1/2$ and
hence \eqref{smallnorm} holds true with a suitable $C_2$ which
depends only on the single site potential, the single site
distribution, and the function $\chi$. Moreover, by construction
$\{ \tilde{\phi}_1, \ldots , \tilde{\phi}_j \}$ form an
orthonormal set and obey \eqref{ortho}. This concludes the proof
of the lemma.
\end{proof}
We are now in position to give the
\begin{proof}[Proof of Theorem \ref{wegner}]
We closely follow the proof of Theorem 4.1 in \cite{CKM} and make
the necessary modifications. Let $\beta,\sigma,I$ be as above and
for each odd $L\in \N$, we set $n_L = \lfloor \tau (L/2)^\beta
\rfloor + 1$ with some $\tau > 0$ to be chosen later. For every
$\lambda \in I$ and $\theta > 0$, we define the events
\begin{align*}
A_\theta^{(\lambda,L)} & = \{ \| g_\lambda(-(L+1)/2 + n_L ) \cdot
\ldots \cdot g_\lambda(-(L-1)/2) (0,1)^t \| > e^{\theta
(L/2)^\beta} \},\\ B_\theta^{(\lambda,L)} & = \{ \|
g_\lambda((L+1)/2 - n_L)^{-1} \cdot \ldots \cdot
g_\lambda((L-1)/2)^{-1} (0,1)^t \|
> e^{\theta (L/2)^\beta} \}.
\end{align*}
Let $\kappa = \tau \alpha_1/2\delta$ with $\alpha_1$ and $\delta$
from Lemma~\ref{weglem1}. Then
\begin{align*}
\p \{ \dist & (\lambda, \sigma( H_\Lambda(\omega)) \le e^{-\sigma
L^\beta} \} \le \\ \le & \; \p \left\{ \{ \dist (\lambda,\sigma(
H_\Lambda(\omega)) \le e^{-\sigma L^\beta} \} \cap
\bigcap_{|\lambda - \lambda'| \le e^{-\sigma L^\beta}}
\left(A_{\kappa/2}^{(\lambda',L)} \cap
B_{\kappa/2}^{(\lambda',L)}\right) \right\}\\ & + \p \left\{
A_\kappa^{(\lambda,L)} \cap B_\kappa^{(\lambda,L)} \cap
\bigcup_{|\lambda - \lambda'| \le e^{-\sigma L^\beta}}
(A_{\kappa/2}^{(\lambda',L)})^c \right\}\\ & + \p \left\{
A_\kappa^{(\lambda,L)} \cap B_\kappa^{(\lambda,L)} \cap
\bigcup_{|\lambda - \lambda'| \le e^{-\sigma L^\beta}}
(B_{\kappa/2}^{(\lambda',L)})^c \right\}\\ & + \p \{
(A_\kappa^{(\lambda,L)})^c \} + \p \{ (B_\kappa^{(\lambda,L)})^c
\}\\ = & \; {\rm (i) + (ii) + (iii) + (iv)}
\end{align*}
and Lemma \ref{weglem1} immediately implies
\begin{equation}\label{eins}
{\rm (iv)} \le 2 e^{-\frac{1}{2} \tau \alpha_1 (L/2)^\beta}
\end{equation}
provided $L$ is large enough. Together with Lemma~\ref{L2bound},
Lemma~\ref{weglem2} yields for $L$ large enough,

\begin{equation}\label{zwei}
{\rm (i)} \le \tilde{C} L \max \left\{ e^{-\sigma L^\beta \rho} ,
e^{-\frac{1}{2} \kappa (L/2)^\beta \rho} \right\}
\end{equation}
with $\tilde{C}$ independent of $\lambda$ and $L$.

Finally, using \eqref{gron2}, for $L$ large enough, one proves similarly to \cite{CKM}

\begin{equation}\label{drei}
{\rm (ii) + (iii)} \le 2 e^{-\alpha_2 (L/2)^\beta}
\end{equation}
for some suitable $\alpha_2 > 0$ if $\tau$ is chosen small enough.
The assertion now follows from \eqref{eins}---\eqref{drei}.
\end{proof}

\setcounter{equation}{0}


\section{The Initial Length Scale Estimate}

Fix $\lambda \in \R \setminus M$ and let $\nu_\lambda$ be the
unique $\mu_\lambda$-invariant measure on $P(\R^2)$, see
(\ref{geqn}), and $\delta(\overline{x},\overline{y})$ the
projective distance of $\overline{x}, \overline{y} \in P(\R^2)$.

The measure $\nu_\lambda$ is H\"older continuous:

\begin{Lemma}\label{L6.1}
There exist $\rho > 0$ and $C >0$ such that for all $\overline{x} \in P(\R^2)$ and $\varepsilon > 0$, one has
$$
\nu_\lambda (\{ \overline{y} : \delta(\overline{x},\overline{y}) \le \varepsilon \} ) \le C \varepsilon^\rho.
$$
\end{Lemma}
\begin{proof} This follows from Corollary VI.4.2 of \cite{Bougerol/Lacroix}. Note that in the case $G(\lambda) \subset {\rm SL}(2,\R)$ the assumptions required there are equivalent to $\int \|g\|^\tau d\mu_\lambda(g) < \infty$ for some $\tau > 0$ and that $G(\lambda)$ is non-compact and strongly irreducible. In particular, non-compactness is equivalent to the contractivity required in \cite{Bougerol/Lacroix}; see \cite[Prop. IV.4.11]{Carmona/Lacroix}. Integrability of $\|g\|^\tau$ with respect to $\mu_\lambda$ for all $\tau > 0$ follows from \eqref{gron1} and boundedness of the distribution of $q_n$. \end{proof}

We know that $\frac1n \log \| U_\lambda(n)\|$ converges to $\gamma(\lambda)$ in expectation. We will need a large deviation result for this limit. In fact, the following result on the asymptotics of $\|U_\lambda(n) x\|$ for any initial vector $x \not= 0$ holds.

\begin{Lemma}\label{L6.2}
There exists $\alpha > 0$ such that for every $\varepsilon > 0$ and $x \not=0$, one has
$$
\limsup_{n \rightarrow \infty} \frac1n \log \p ( | \log \|U_\lambda(n)x\| - n \gamma(\lambda)| > n \varepsilon) < \alpha.
$$
\end{Lemma}
\begin{proof} This follows from Theorem V.6.2 of \cite{Bougerol/Lacroix}, whose assumptions, that is, $G(\lambda)$ strongly irreducible and $\int \|g\|^\tau d\mu_\lambda(g) < \infty$ for some $\tau > 0$, are satisfied. \end{proof}

An immediate consequence of Lemma \ref{L6.2} is that for every $\varepsilon > 0$ and $x \not= 0$, there exists $n_0 \in \N$ such that

\begin{equation}\label{6.1}
\p \left( e^{(\gamma(\lambda) - \varepsilon)n} \le \|U_\lambda(n) x\| \le e^{(\gamma(\lambda) + \varepsilon)n} \right) \ge 1 - e^{-\alpha n} \mbox{  for } n \ge n_0.
\end{equation}

Next, we establish a large deviation result for $|\langle U_\lambda(n)x,y \rangle|$, that is, in particular for the matrix elements of the transfer matrices.

\begin{Lemma}\label{L6.3}
Fix $y$ with $\|y\|=1$. For all $\varepsilon > 0$, there are $n_0 \in \N$ and $\delta_0 > 0$ such that
$$
\sup_{x \not= 0} \p \left( \frac{|\langle U_\lambda(n) x , y \rangle|}{\|U_\lambda(n) x\|} < e^{-\varepsilon n} \right) < e^{-\delta_0 n} \mbox{  for } n \ge n_0.
$$
\end{Lemma}
\begin{proof} We closely follow the proof of Prop.~VI.2.2 in \cite{Bougerol/Lacroix}. Define $f_n : [0,1] \rightarrow \R$ by
$$
f_n(t) = \left\{
\begin{array}{cl}
1 & \mbox{ if } 0 \le t \le e^{-\varepsilon n},\\
2 - t e^{\varepsilon n} & \mbox{ if } e^{-\varepsilon n} \le t \le 2 e^{-\varepsilon n},\\
0 & \mbox{ if } 2 e^{-\varepsilon n} \le t \le 1.
\end{array}
\right.
$$

Then

\begin{equation}\label{6.2}
|f_n(t) - f_n(t')| \le |t-t'| e^{\varepsilon n} \mbox{ for all } t,t' \in [0,1].
\end{equation}

Define $\Phi_n : P(\R^2) \rightarrow \R$ by
$$
\Phi_n(\overline{z}) = f_n \left( | \langle \tfrac{z}{\|z\|} , y \rangle | \right).
$$

For $\overline{z},\overline{z}' \in P(\R^2)$, choose representatives $z$ and $z'$ such that $\|z\| = \|z'\| = 1$ and the angle between $z$ and $z'$ is at most $\frac{\pi}{2}$. Then, by \eqref{6.2},

\begin{equation}\label{6.3}
| \Phi_n(\overline{z}) - \Phi_n(\overline{z}') |  \le | | \langle z,y \rangle | - | \langle z',y \rangle | | e^{\varepsilon n} \le \|z - z'\| e^{\varepsilon n} \le \sqrt{2} \delta(\overline{z},\overline{z}') e^{\varepsilon n},
\end{equation}
recall that $\delta(\overline{z},\overline{z}') = |\sin ( \mbox{angle}(z,z') )|$. This implies, using notation from Section 3,

\begin{equation}\label{6.4}
\|\Phi_n\|_\alpha = \|\Phi_n\|_\infty + m_\alpha(\Phi_n) \le \sqrt{2} 1 + e^{\varepsilon n} \mbox{ for all } 0 < \alpha < 1.
\end{equation}

The definition of $\Phi_n$ shows

\begin{align}\label{6.5}
\p \left( \frac{|\langle U_\lambda(n) x , y \rangle|}{\|U_\lambda(n) x\|} < e^{-\varepsilon n} \right) & \le \E (\Phi_n(U_\lambda(n) \overline{x}))\\ \nonumber
& \le \left| \E (\Phi_n(U_\lambda(n) \overline{x})) - \int \Phi_n d\nu_\lambda \right| + \int \Phi_n d\nu_\lambda.
\end{align}

Noting that $\E(\Phi_n(U_\lambda(n) \overline{x})) = (T_\lambda^n \Phi_n)(\overline{x})$, we may use Lemma \ref{L3.7} above to conclude that for $0 < \alpha \le \alpha_0$, there exists $\rho < 1$ and $n_0 \in \N$ such that for $n \ge n_0$ and all $x$,

\begin{equation}\label{6.6}
\left\| \E (\Phi_n(U_\lambda(n) \overline{x})) - \int \Phi_n d\nu_\lambda \right\| \le C_\alpha \rho^n \| \Phi_n\|_\alpha \le \rho^n ( \sqrt{2} e^{\varepsilon n} + 1) \le \tfrac12 e^{-\delta_1 n},
\end{equation}
where in the last step, $\varepsilon$ is assumed to be sufficiently small, that is, such that $\log \rho + \varepsilon < 0$, and $\delta_1 := \frac12 | \log \rho + \varepsilon |$.

Next choose the unit vector $w = (w_1,w_2)^t = (y_2,-y_1)^t$. We therefore have $| \langle u/\|u\| , y \rangle | = \delta(\overline{u},\overline{w})$ for all $u \not= 0$. It then follows from Lemma \ref{L6.1} that there exist $\beta > 0$ and $C > 0$ such that

\begin{align}\label{6.7}
\int \Phi_n d\nu_\lambda & \le \nu_\lambda \{\overline{u} : | \langle u/\|u\|,y \rangle | \le 2 e^{-\varepsilon n} \} \\ \nonumber
& = \nu_\lambda \{ \overline{u} : \delta(\overline{u},\overline{w}) \le 2 e^{-\varepsilon n} \} \le 2^\beta C e^{-\varepsilon \beta n}.
\end{align}

Inserting \eqref{6.6} and \eqref{6.7} into \eqref{6.5} completes the proof of Lemma \ref{L6.3} with $0 < \delta_0 < \min \{ \delta_1,\varepsilon \beta\}$. We have assumed that $\varepsilon$ is sufficiently small, but the result extends to large $\varepsilon$ with unchanged $\delta_0$. \end{proof}

Note that results corresponding to Lemma \ref{L6.2} and Lemma
\ref{L6.3} also hold for $n \rightarrow - \infty$ since the result
from \cite{Bougerol/Lacroix} can be applied in the same way to the
products of random matrices $g_\lambda^{-1}(n) \times \cdots
\times g_\lambda^{-1}(-1)$ for $n < 0$ and the Lyapunov exponent
can equivalently be defined as
\begin{equation*}
\gamma(\lambda) = \lim_{n \rightarrow - \infty} \frac{1}{|n|} \E
\left( \M g_{\lambda}^{-1}(n) \cdot ... \cdot g_{\lambda}^{-1}(-1)
\M \right).
\end{equation*}

We can now combine Lemma \ref{L6.2} and Lemma \ref{L6.3} to show that with large probability, the matrix elements $| \langle U_\lambda(n)x,y \rangle |$ grow exponentially at the rate of almost the Lyapunov exponent.

\begin{Corollary}\label{C6.4}
Let $\|x\|= \|y\|= 1$. Then for every $\varepsilon > 0$, there exist $\delta > 0$ and $n_0 \in \N$ such that
$$
\p \left( | \langle U_\lambda(n) x,y \rangle | \ge e^{(\gamma(\lambda) - \varepsilon)n} \right) \ge 1 - e^{-\delta n} \mbox{ for } n \ge n_0.
$$
\end{Corollary}
\begin{proof}
From \eqref{6.1} and Lemma \ref{L6.3} we get for all $\varepsilon
> 0$, $$ | \langle U_\lambda(n) x,y \rangle | \ge e^{-\varepsilon
n} \|U_\lambda(n) x\| \ge e^{(\gamma(\lambda) - 2\varepsilon)n}
\mbox{ for } n \ge n_0 $$ with probability at least $1 -
e^{-\alpha n} -e^{-\delta_0 n}$. This yields the assertion.
\end{proof}

We are now ready to state and prove the main result of this section, an initial length scale estimate at energy $\lambda$. Let $L \in 3\Z \setminus 6\Z$ and $\Lambda = \Lambda_L(0) = [-L/2,L/2]$. Let $H_\Lambda(\omega)$ be the restriction of $H_\omega$ to $\Lambda$ with Dirichlet boundary conditions at $-L/2$ and $L/2$. For $\lambda \not\in \sigma(H_\Lambda(\omega))$, let $R_\Lambda(\lambda) = (H_\Lambda(\omega) - \lambda)^{-1}$. Define the characteristic functions $\chi^{{\rm int}} = \chi_{\Lambda_{L/3}(0)} = \chi_{[-L/6,L/6]}$ and $\chi^{{\rm out}} = \chi_{\Lambda_L(0) \setminus \Lambda_{L-2}(0)} = \chi_{[-L/2,-L/2+1] \cup [L/2-1,L/2]}$.

For $\gamma, \lambda \in \R$ and $\omega \in \Omega$, let us call the cube $\Lambda$ $(\gamma,\lambda)$-good for $\omega$ if $\lambda \not\in \sigma(H_\Lambda(\omega))$ and $\| \chi^{{\rm out}} R_\Lambda(\lambda) \chi^{{\rm int}} \| \le e^{-\gamma L/3}$. We have the following theorem:

\begin{Theorem}\label{T6.5}
For every $\varepsilon > 0$, there exist $\delta > 0$ and $L_0 \in \N$ such that for $L \ge L_0$ with $L \in 3\Z \setminus 6\Z$, we have

\begin{equation}\label{feilse}
\p \left\{  \Lambda \mbox{ is  $(\gamma(\lambda) - \varepsilon , \lambda)$-good for $\omega$ }\right\} \ge 1 - e^{-\delta L}.
\end{equation}
\end{Theorem}
\begin{proof}
Let $u_\pm$ be the solutions of $H_\omega u = \lambda u$ with Dirichlet boundary conditions at $\pm L/2$, that is, $u_+(L/2) = u_-(-L/2) = 0$, $u_+'(L/2) = u_-'(-L/2) = 1$. Then the Green's function $G_\Lambda(\lambda,x,y)$ (i.e., the kernel of $R_\Lambda(\lambda)$) is given by

\begin{equation}\label{6.8}
G_\Lambda(\lambda,x,y) = \frac{1}{W(u_+,u_-)} \left\{ \begin{array}{cl} u_+(x) u_-(y) & \mbox{ for } x \ge y\\u_-(x) u_+(y) & \mbox{ for } x < y \end{array} \right. ,
\end{equation}
where the Wronskian $$ W(u_+,u_-)(x) = u_+(x) u_-'(x) -
u_+'(x)u_-(x) $$ is constant in $x$. Setting $x = L/2$, we get $$
W(u_+,u_-) = \langle (u_-(L/2),u_-'(L/2))^t, (-1,0)^t \rangle =
\langle A_{\lambda}(L/2,-L/2) (0,1)^t, (-1,0)^t \rangle, $$ where
$A_{\lambda}(x,y)$ denotes the transfer matrix from $y$ to $x$. By
stationarity we can use Corollary \ref{C6.4} to conclude

\begin{equation}\label{6.9}
\p \left( |W(u_+,u_-)| \ge e^{(\gamma(\lambda) - \varepsilon)L} \right) \ge 1 - e^{-\delta L} \mbox{ for } L \ge L_0.
\end{equation}

Note that $\lambda \in \sigma(H_\Lambda(\omega))$ if and only if $W(u_+,u_-) = 0$. Thus the event in \eqref{6.9} implies $\lambda \not\in \sigma(H_\Lambda(\omega))$.

Let $x \in [L/2-1,L/2]$ and $y \in [-L/6,L/6]$. Then

\begin{equation}\label{6.10}
|u_+(x)| \le C \mbox{ uniformly in } \omega \mbox{ and } L
\end{equation}
by Lemma A.1. Also

\begin{align*}
|u_-(y)| & \le \| ( u_-(y) , u_-'(y) )^t \| = \|
A_{\lambda}(y,-L/2) (0,1)^t \| \\ & \le \| A_{\lambda}(y,\lfloor y
+ 1/2 \rfloor - 1/2) \| \cdot \| A_{\lambda}(\lfloor y + 1/2
\rfloor -1/2, -L/2) (0,1)^t \|.
\end{align*}
Again by Lemma A.1,

\begin{equation}\label{6.11}
\|A_{\lambda}(y,\lfloor y +1/2 \rfloor - 1/2)\| \le C,
\end{equation}
and by stationarity and \eqref{6.1} (note $-L/6 \le \lfloor y +
1/2 \rfloor - 1/2 \le L/6$),

\begin{equation}\label{6.12}
\p \left( \|A_{\lambda} (\lfloor y + 1/2 \rfloor - 1/2, -L/2)
(0,1)^t\| \le e^{(\gamma(\lambda) + \varepsilon) 2L/3} \right) \ge
1 - e^{-\alpha L/3} \mbox{ if } L/3 \ge L_0.
\end{equation}

Combining \eqref{6.9}---\eqref{6.12} and using
$G_\Lambda(\lambda,x,y) = u_-(y)u_+(x)/W(u_+,u_-)$, we get $$ \p
\left( |G_\lambda(\lambda,x,y)| \le C e^{-(\gamma(\lambda) - 5
\varepsilon)L/3} \right) \ge 1 - e^{-\delta L} - e^{-\alpha L/3}
$$ for $L$ sufficiently large. In a completely analogous way the
same estimate is found if $x \in [-L/2,-L/2+1]$, $y \in
[-L/6,L/6]$. From this it can now be seen easily that for every
$\varepsilon > 0$, there exist $\delta > 0$ and $L_0 \in \N$ such
that for $L \ge L_0$ and $L \in 3\Z \setminus 6\Z$, we have $$
\sup_x \int | \chi^{{\rm out}}(x) G_\Lambda(\lambda,x,y)
\chi^{{\rm int}}(y)| dy \le e^{- (\gamma(\lambda) -
\varepsilon)L/3} $$ and $$ \sup_y \int | \chi^{{\rm out}}(x)
G_\Lambda(\lambda,x,y) \chi^{{\rm int}}(y)| dx \le e^{-
(\gamma(\lambda) - \varepsilon)L/3} $$ with probability at least
$1 - e^{-\delta L}$. The theorem now follows by Schur's test.
\end{proof}

Theorem~\ref{T6.5} can be regarded as a fixed-energy initial length scale estimate. Our goal is to start a variable-energy multiscale induction which will ultimately yield both Theorem~\ref{exploc} and Theorem~\ref{dynloc}. We therefore need an estimate of the form \eqref{feilse} where the energy is not fixed, but rather varying over an interval. A result of this kind will be established, using the Wegner estimate and an argument from \cite{VDK}, in the following corollary.

\begin{Corollary}\label{C6.6}
For every $\lambda \in I$ and every $\beta \in (0,1)$, $\sigma,
\varepsilon > 0$, let $\alpha>0,\delta > 0$ and $L_0 \in \N$ be as
in Theorem~\ref{wegner} and Theorem~\ref{T6.5}, respectively. For
every $0 < \varepsilon < \varepsilon'$ and every $L \ge L_0$, let
$$ \kappa_L = \frac{1}{2} e^{-2\sigma L^\beta} \left( e^{-
(\gamma(\lambda) - \varepsilon')L/3} - e^{- (\gamma(\lambda) -
\varepsilon)L/3} \right). $$ Then we have for every $L \ge L_0$

\begin{equation}\label{aveilse}
\p \left\{ \forall \lambda' \in (\lambda - \kappa_L , \lambda + \kappa_L): \, \Lambda \mbox{ is  $(\gamma(\lambda) - \varepsilon' , \lambda')$-good} \right\} \ge 1 - e^{-\delta L} - e^{-\alpha L^\beta}.
\end{equation}
\end{Corollary}

\begin{proof}
With probability $1 - e^{-\delta L} - e^{-\alpha L^\beta}$ we have that both the event in \eqref{feilse} and the complementary event in \eqref{weges} hold. Thus, by assumption we have that for every $\lambda' \in (\lambda - \kappa_L , \lambda + \kappa_L)$, we have $\lambda' \not\in \sigma(H_\Lambda(\omega))$ and moreover, by the resolvent equation,

\begin{align*}
\| \chi^{{\rm out}} R_\Lambda(\lambda') \chi^{{\rm int}} \| = & \| \chi^{{\rm out}} \left( R_\Lambda(\lambda) + (\lambda - \lambda') R_\Lambda(\lambda') R_\Lambda(\lambda) \right) \chi^{{\rm int}} \| \\
\le & \, \| \chi^{{\rm out}} R_\Lambda(\lambda) \chi^{{\rm int}} \| + | \lambda - \lambda' | \cdot \| \chi^{{\rm out}} R_\Lambda(\lambda') R_\Lambda(\lambda) \chi^{{\rm int}} \| \\
\le & \, \| \chi^{{\rm out}} R_\Lambda(\lambda) \chi^{{\rm int}} \| + | \lambda - \lambda' | \cdot \| R_\Lambda(\lambda')\| \cdot \| R_\Lambda(\lambda)\| \\
\le & \, e^{- (\gamma(\lambda) - \varepsilon)L/3} + 2 \cdot | \lambda - \lambda' | \cdot e^{2 \sigma L^\beta} \\
\le & \, e^{- (\gamma(\lambda) - \varepsilon')L/3}.
\end{align*}
Thus for these $\omega$'s, the cube $\Lambda$ is $(\gamma(\lambda) - \varepsilon' , \lambda')$-good.
\end{proof}

\setcounter{equation}{0}


\section{Proof of the Main Theorems}

In the preceding two sections we have established the two
ingredients, namely, a Wegner estimate and an initial length scale
estimate, that are necessary to start the multiscale induction
which, by known results, implies both Theorem~\ref{exploc} and
Theorem~\ref{dynloc}. In this section we briefly show how to
reduce these two theorems to known results, given
Theorem~\ref{wegner} and Corollary~\ref{C6.6} above.

Let $M$ be the discrete set found in Theorem \ref{gamma+}.

\begin{proof}[Proof of Theorem~\ref{exploc}.]
Fix an arbitrary compact interval $I \subset \R \setminus M$. It
follows from Theorem~\ref{wegner} and Theorem~\ref{T6.5} that we
have both a Wegner estimate and a fixed energy initial length
scale estimate for every $\lambda \in I$. Corollary~\ref{C6.6}
shows that these two results imply a variable-energy initial
length scale estimate for a ball $B(\lambda)$ of explicit radius
around $\lambda$ . The variable-energy multiscale analysis as
presented, for example, in \cite{Stollmann} then establishes
variable-energy resolvent decay estimates on a sequence $(L_k)_{k
\in \N}$ of length scales for energies in $B(\lambda)$. These
estimates, together with the existence of polynomially bounded
eigenfunctions for spectrally almost every energy, yield pure
point spectrum in $B(\lambda)$ with exponentially decaying
eigenfunctions for almost every $\omega \in \Omega$; see, for
example, \cite{Stollmann} for details. Thus, we have exponential
localization in $\R \setminus M$ for almost every $\omega \in
\Omega$. Finally, since, by general principles
\cite{Carmona/Lacroix}, the set $M$ carries almost surely no
spectral measure, we have exponential localization in $\R$ for
almost every $\omega \in \Omega$ and hence Theorem~\ref{exploc}.
\end{proof}

\begin{proof}[Proof of Theorem~\ref{dynloc}.]
It essentially follows from \cite{DS} that the variable-energy
resolvent decay estimates, as given by the output of the the
variable-energy multiscale analysis, imply strong dynamical
localization in the sense of Theorem~\ref{dynloc}. For the curious
reader we briefly sketch the argument, referring him to \cite{DS}
for necessary notation. Given a compact interval $I \subset \R
\setminus M$, a compact set $K \subset \R$, and $p
> 0$, we first let $\gamma = \min \{ \gamma(\lambda) : \lambda \in I\} > 0$. Next we choose $L_1$ large enough so that, for every
$\lambda \in I$, Theorem~\ref{wegner} and Corollary~\ref{C6.6}
imply both $W(I,L,\Theta,q)$, $L \ge L_1$, and $G(B(\lambda), L_1,
\gamma - \varepsilon', \xi)$ of \cite{DS} with parameters
sufficient to cover the desired $p$. Having this length scale
fixed, we decompose the interval $I$ into a finite disjoint union
of intervals $I_1,\ldots,I_m$, each of them having length bounded
by $\kappa_{L_1}$. We split the projection $P_I(H_\omega)$ in
\eqref{sdl} into the finite sum $\sum_{i = 1}^m P_{I_i}(H_\omega)$
and treat each term separately. For every $i$, we can apply
Theorem~3.1 of \cite{DS}, with initial length scale $L_1$, since
all the other conditions (e.g., (INDY), (GRI), (WEYL), (EDI), (i))
are known to hold for the concrete operators $H_\omega$ under
consideration \cite{Stollmann}. After establishing \eqref{sdl},
with $I$ replaced by $I_i$, for every $i$, we get \eqref{sdl} and
hence Theorem~\ref{dynloc}
\end{proof}

\setcounter{equation}{0}


\begin{appendix}
\section{A Priori Solution Estimates}

Here we provide several a priori estimates for solutions of the
Schr\"odinger equation which are used repeatedly in the main text.
For convenience we include proofs of these standard facts.

\begin{Lemma} \label{growthbound}
Let $q\in L^1_{{\rm loc}}(\R)$, $u$ a solution of $-u''+qu=0$ and $x,y
\in \R$. Then

\begin{equation*}
|u(x)|^2+|u'(x)|^2 \le \left( |u(y)|^2+|u'(y)|^2 \right) \exp
\left\{ \int_{\min(x,y)}^{\max(x,y)} |q(t)+1|\,dt \right\}.
\end{equation*}
\end{Lemma}

\begin{proof}
For $R(t) := |u(t)|^2+|u'(t)|^2$ one has $$|R'(t)| = |2(q(t)+1)
{\rm Re} u'(t)\overline{u(t)}| \le |q(t)+1| R(t),$$ i.e.\ $|(\ln
R(t))'| \le |q(t)+1|$, which implies the lemma. \end{proof}

\begin{Lemma} \label{diffbound}
For $i=1,2$, let $q_i \in L^1_{{\rm loc}}(\R)$, $u_i$ solutions of
$-u_i''+q_iu_i=0$ with $u_1(y)=u_2(y)$ and $u_1'(y)=u_2'(y)$ for
some $y\in \R$. Then for any $x\in \R$,

\begin{eqnarray*}
\lefteqn{\left( |u_1(x)-u_2(x)|^2 + |u_1'(x)-u_2'(x)|^2
\right)^{1/2} \le}
\\
& & \left( |u_1(y)|^2+|u_1'(y)|^2 \right)^{1/2} \exp \left\{
\int_{\min(x,y)}^{\max(x,y)} (|q_1(t)|+|q_2(t)|+2)\,dt \right\}
\times \\ & & \times \int_{\min(x,y)}^{\max(x,y)}
|q_1(t)-q_2(t)|\,dt.
\end{eqnarray*}
\end{Lemma}

\begin{proof}
Without restriction let $y\le x$. The solutions $u_1$ and $u_2$
satisfy
\begin{eqnarray*}
\left( \begin{array}{c} u_1(x)-u_2(x) \\ u_1'(x)-u_2'(x)
\end{array} \right) & = & \int_y^x \left( \begin{array}{c} 0 \\
(q_1(t)-q_2(t)) u_1(t) \end{array} \right) \,dt + \\ & & \mbox{}+
\int_y^x \left(
\begin{array}{cc} 0 & 1 \\ q_2(t) & 0 \end{array} \right) \left(
\begin{array}{c} u_1(t)-u_2(t) \\ u_1'(t)-u_2'(t) \end{array} \right)\,dt,
\end{eqnarray*}
and thus
\begin{eqnarray} \label{A1}
\left\| \left( \begin{array}{c} u_1(x)-u_2(x) \\ u_1'(x)-u_2'(x)
\end{array} \right) \right\| & \le & \int_y^x
|q_1(t)-q_2(t)||u_1(t)|\,dt + \\ & & \mbox{}+ \int_y^x
(|q_2(t)|+1) \left\| \left(
\begin{array}{c} u_1(t)-u_2(t) \\ u_1'(t)-u_2'(t) \end{array} \right)
\right\|\,dt. \nonumber
\end{eqnarray}
Gronwall's lemma, e.g.\ \cite{Walter}, yields $$\left\| \left(
\begin{array}{c} u_1(x)-u_2(x) \\ u_1'(x)-u_2'(x)
\end{array} \right) \right\| \le \int_y^x |q_1(t)-q_2(t)||u_1(t)|\,dt
\exp \left\{ \int_y^x (|q_2(t)|+1)\,dt \right\}.$$ By
Lemma~\ref{growthbound} we have for all $t\in[y,x]$ that $$
|u_1(t)| \le \exp \left\{ \frac{1}{2} \int_y^x |q_1(s)+1|\,ds
\right\} \left( |u_1(y)|^2 + |u_1'(y)|^2 \right)^{1/2}.$$
Inserting this into (\ref{A1}) yields the result. \end{proof}

\begin{Lemma} \label{L2bound}
Let $q\in L^1_{{\rm loc,unif}}(\R)$, i.e.\ $\|q\|_{1,{\rm unif}} := \sup_x
\int_x^{x+1} |q(t)|\,dt < \infty$. Then there is a $C>0$, only
depending on $\|q\|_{1,{\rm unif}}$ such that for all solutions $u$ of
$-u''+qu=0$ and all $x\in\R$
\begin{equation} \label{A2}
\int_{x-1}^{x+1} |u(t)|^2\,dt \ge C \left( |u(x)|^2 + |u'(x)|^2
\right). \end{equation} \end{Lemma}

\begin{proof}
By Lemma~\ref{growthbound} there are constants $0<C_1, C_2<\infty$
only depending on $\|q\|_{1,{\rm unif}}$ such that for all $t\in
[x-1,x+1]$ $$C_1 (|u(x)|^2 + |u'(x)|^2) \le |u(t)|^2 + |u'(t)|^2
\le C_2 (|u(x)|^2 + |u'(x)|^2).$$ With $C_3 := (C_1/2)^{1/2}$ and
$C_4 := (2C_2)^{1/2}$ we get $$C_3 (|u(x)|+|u'(x)|) \le |u(t)| +
|u'(t)| \le C_4 (|u(x)|+|u'(x)|).$$ It now follows from elementary
geometric considerations, e.g.\ \cite[p.218]{Boundsol}, that
$[x-1,x+1]$ contains an interval of length $\min(2,C_3/4C_4)$ on
which $|u| \ge C_3(|u(x)|+|u'(x)|)/4$. This yields (\ref{A2}).
\end{proof}

\setcounter{equation}{0}


\section{Cocycles and invariant measures}

In this appendix we collect some basic facts about cocycles and
the existence and uniqueness of invariant measures for group
actions, which are used in Section 3 to prove H\"older continuity
of the Lyapunov exponent. All this can be found in Chapter IV of
\cite{Carmona/Lacroix}.

Let $G$ be a metric group, with unit $e$, that is both locally
compact and $\sigma$-compact. Let $B$ be a metrizable topological
space such that $G$ acts on $B$, i.e. to each $(g,b) \in G \times
B$ one can continuously associate an element $gb \in B$ for which
$(g_1g_2) \cdot b = g_1 \cdot (g_2 \cdot b)$ for $g_1,$ $g_2 \in
G$ and $b \in B$, and $e \cdot b = b$ for $b \in B$. A continuous
map $\sigma:G \times B \rightarrow (0, \infty)$ is called a
cocycle if for all $g_1,g_2 \in G$ and $b \in B$ one has
$\sigma(g_1g_2,b)=\sigma(g_1,g_2b)\sigma(g_2,b).$

Note that if $\sigma$ is a cocycle, then clearly $\sigma^t$ is
also a cocycle for all $t \in \R$. There is one particular example
of primary importance:

\begin{Example} \label{ex}
Let $B$ be a compact metric space, and $G$ be a group acting on
$B$. Take $\tilde{B}: = B \times B \setminus \{ (b,b): b \in B
\}$, and consider the induced action of $G$ on $\tilde{B}$, i.e.
$g \cdot (a,b):=(ga,gb)$ for all $g \in G$ and $(a,b) \in
\tilde{B}$. If $\delta$ is the metric on $B$, then
\begin{equation*}
\sigma_1 \left(g, (a,b) \right) := \frac{ \delta(ga,gb)}{
\delta(a,b)},
\end{equation*}
defines a cocycle on $G \times \tilde{B}$.
\end{Example}

A crucial property of cocycles is that they satisfy certain
integral estimates. For this reason, the ``pseudo-convolution'' of
a probability measure $\mu$ on $G$ and a measure $\nu$ on $B$ is
introduced by:
\begin{equation*}
( \mu * \nu)(f) := \int f(gb) d \mu(g) d \nu(b),
\end{equation*}
for all $f \in \Cal{B}(B)$, the bounded measurable functions on
$B$. Here if $B$=$G$, then $\mu^n$ is $\mu * \mu * . . . * \mu$,
$n$-times, and the above definition coincides with that of the
ordinary convolution on $G$. We note that a cocycle is said to be
$\mu$-integrable if $ \overline{ \sigma}(g) := \sup_{b \in B}
\sigma(g,b)$ is $\mu$-integrable.

The concept of invariance will also be important. A measure $\nu$
on $B$ is said to be $\mu$-invariant if $\mu * \nu = \nu$. For $B$
compact, the existence of an invariant measure is trivial. In
particular, any weak limit of the sequence $\frac{1}{n}
\sum_{j=1}^n \mu^n * m$, where $m$ is an arbitrary probability
measure on $B$, is a $\mu$-invariant probability measure. In
addition an operator $T: \Cal{B}(B) \rightarrow \Cal{B}(B)$ is
defined by:
\begin{equation*}
(Tf)(b):= \int f(gb)d \mu(g).
\end{equation*}
Note that the operator $T^n$ is given by the formula above with
$\mu^n$ replacing $\mu$. The relationship between these operators
and $\mu$-invariant probability measures is illustrated in Lemma
$\ref{cocycle}$ below.

Lastly, if $B$ is a compact space metric space, with metric
$\delta$, then for any real number $\alpha$ the space of $\alpha$-H\"{o}lder continuous functions, $\Cal{L}_{\alpha}(B)$, is defined
to be
\begin{equation*}
\Cal{L}_{\alpha}(B):= \{ f \in C(B): m_{\alpha}(f) < \infty \},
\end{equation*}
where
\begin{equation*}
m_{\alpha}(f) : = \sup_{(a,b) \in \tilde{B}} \frac{|f(a)-f(b)|}{
\delta^{\alpha}(a,b)}.
\end{equation*}
Equipped with the norm
\begin{equation*}
\M f \M_{\alpha} = \M f \M_{\infty} + m_{\alpha}(f),
\end{equation*}
$\Cal{L}_{\alpha}(B)$ is a Banach space.

\begin{Lemma} \label{cocycle}
Consider the cocycle $\sigma_1$, as defined in Example $\ref{ex}$,
on $G \times \tilde{B}$. Suppose

\noindent {\rm (i)} there exists a positive number $\tau$ such that
$\sigma_1^t$ is $\mu$- integrable for $|t| \leq \tau$, and

\noindent {\rm (ii)} there exists an integer $N$ such that
\begin{equation*}
\sup_{b \in \tilde{B}} \int \log [ \sigma_1(g,b) ] d \mu^N(g) < 0.
\end{equation*}
Then, there exists a real number $\alpha_0$ such that for any
$\alpha$ with $0< \alpha \leq \alpha_0$ there exists constants
$C_{\alpha} < \infty$ and $\rho_{\alpha}<1$ for which:

\noindent {\rm (a)}
\begin{equation*}
\sup_{b \in B} \int \sigma_1(g,b)^{\alpha}d \mu^n(g) \leq
C_{\alpha} \rho_{\alpha}^n
\end{equation*}
for $n=1,2,...$.

\noindent {\rm (b)} $T$ is a bounded operator on $\Cal{L}_{\alpha}$
satisfying
\begin{equation*}
\M T^nf- \nu(f) \M_{\alpha} \leq \M f \M_{\alpha} C_{\alpha}
\rho_{\alpha}^n
\end{equation*}
for $n=1,2,...$ and $f \in \Cal{L}_{\alpha}$, where $\nu$ is any
$\mu$-invariant probability measure. In particular, this proves
uniqueness of the invariant measure.

\noindent {\rm (c)} The operator $T$ on $\Cal{L}_{\alpha}$ has
eigenvalue 1 and the rest of the spectrum is contained in a disk
of radius strictly less than 1. Moreover, $T$ admits the following
decomposition:
\begin{equation*}
T^nf = \nu(f) + Q^nf \ \ \mbox{for } f \in \Cal{L}_{\alpha},
\end{equation*}
where $\nu$ is the invariant probability measure and $Q$ is an
operator on $\Cal{L}_{\alpha}$ of spectral radius strictly less
than 1.
\end{Lemma}
\begin{proof} See Proposition IV.3.5, Proposition IV.3.15, and Corollary IV.3.16 of \cite{Carmona/Lacroix}.
\end{proof}

\end{appendix}


\baselineskip=12pt
\begin{thebibliography}{99}

\bibitem{Aizenman/Molchanov} M.\ Aizenman and S.\ Molchanov,
Localization at large disorder and at extreme energies: an
elementary derivation. \textit{Commun.\ Math.\ Phys.} {\bf 157}
(1993), 245--278

\bibitem{Aizen+} M.\ Aizenman, J.\ Schenker, R.\ Friedrich, and D.\ Hundertmark,
Finite-volume Criteria for Anderson localization. Preprint 1999, math-ph/9910022v2,
to appear in \textit{Commun.\ Math.\ Phys.}

\bibitem{Avron/Simon} J.Avron and B. Simon, Almost periodic
Schr\"odinger operators, II. The integrated density of states.
\textit{Duke Math. J.} {\bf 50}, (1983), 369--391

\bibitem{BFM} J. Barbaroux, W. Fischer, and P. M\"{u}ller, Dynamical properties of
random Schr\"{o}dinger operators. Preprint 1999, math-ph/9907002

\bibitem{Benderskii/Pastur} M.\ Benderskii and L.\ Pastur, On the
asymptotics of the solutions of second-order equations with random
coefficients (in Russian). \textit{Teoria Funkcii, Func.\ Anal.\ i
Priloz.} (Kharkov University) {\bf N 22} (1975), 3--14

\bibitem{Bi/Ger2} S.\ De Bi\`{e}vre and F.\ Germinet, Dynamical Localization for the
Random Dimer Schr\"{o}dinger Operator. \textit{J.\ Stat.\ Phys.} {\bf 98} (2000), 1135--1148

\bibitem{Borg} G. Borg, Uniqueness theorems in the spectral theory
of $y^{\prime \prime} + ( \lambda - q(x))y=0$. \textit{Proc. 11th
Scandinavian Congress of Mathematicians}, Johan Grundt Tanums
Forlag, Oslo (1952) 276--287

\bibitem{Bougerol/Lacroix} P.\ Bougerol and J.\ Lacroix, \textit{Products of random matrices
with applications to Schr\"odinger operators}. Birkh\"auser, Boston--Stuttgart (1985)

\bibitem{Camp+} M.\ Campanino and A.\ Klein, A supersymmetric transfer matrix and
differentiability of the density of states in the one-dimensional Anderson model.
\textit{Commun.\ Math.\ Phys.} {\bf 104} (1986), 227--241

\bibitem{Carmona1} R.\ Carmona, One-dimensional Schr\"odinger
operators with random potential. \textit{Physica A} {\bf 124}
(1984), 181--188

\bibitem{CLN} R.\ Carmona, Random Schr\"{o}dinger Operators. \textit{\'{E}cole
d'\'{e}t\'{e} de probabilit\'{e}s de Saint-Flour}, XIV---1984, 1--124, Lecture
Notes in Math.\ {\bf 1180}, Springer, Berlin-New York, 1986

\bibitem{CKM} R.\ Carmona, A.\ Klein, and F.\ Martinelli, Anderson localization for
Bernoulli and other singular potentials. \textit{Commun.\ Math.\ Phys.} {\bf 108} (1987), 41--66

\bibitem{Carmona/Lacroix} R. Carmona and J. Lacroix, \textit{Spectral theory of random Schr\"odinger operators}. Birkh\"auser, Basel--Berlin (1990)

\bibitem{Codd/Lev} E.A. Coddington and N. Levinson: Theory of
ordinary differential equations. McGraw-Hill, New York 1955

\bibitem{Combes/Hislop} J.\ M.\ Combes and P.\ D.\ Hislop:
Localization for some continuous random Hamiltonians in
$d$-dimensions. \textit{J.\ Funct.\ Anal.} {\bf 124} (1994),
149--180

\bibitem{DS} D.\ Damanik and P.\ Stollmann, Multi-scale analysis implies strong dynamical
localization. Preprint 1999, mp-arc/99-461, to appear in \textit{Geom.\ Funct.\ Anal.}

\bibitem{Datta/Kundu} P.\ K.\ Datta and K.\ Kundu, The absence of localization in
one-dimensional disordered harmonic chains. \textit{J.\ Phys.\
Condens.\ Matter} {\bf 6} (1994), 4465--4478.


\bibitem{4author} R.\ del Rio, S.\ Jitomirskaya, Y.\ Last, and B.\ Simon, Operators with
singular continuous spectrum, IV. Hausdorf dimension, rank-one perturbations, and
localization. \textit{J.\ Anal.\ Math.} {\bf 69} (1996), 153--200

\bibitem{VDK} H.\ von Dreifus and A.\ Klein, A new proof of localization in the
Anderson tight binding Model. \textit{Commun.\ Math.\ Phys.} {\bf 124} (1989), 285--299

\bibitem{Eastham} M.S.P. Eastham: The Spectral Theory of Periodic
Differential Equations. Scottish Academic Press, Edinburgh--London 1973

\bibitem{Figotin} A.\ Figotin, On the exponential growth of
solutions of finite difference equations with random coefficients
(in Russian). \textit{Dokl.\ Akad.\ Nauk Uzb.\ SSR} 1980(2), 9--11

\bibitem{FK} A.\ Figotin and A.\ Klein, Localization of classical waves. I. Acoustic
waves. \textit{Comm.\ Math.\ Phys.} {\bf 180} (1996), 439--482

\bibitem{Bi/Ger1} F.\ Germinet and S.\ De Bi\`{e}vre, Dynamical localization for discrete
and continuous random Schr\"{o}dinger operators. \textit{Commun.\ Math.\ Phys.} {\bf 194}
(1998), 323--341

\bibitem{Gesztesy/Simon} F. Gesztesy and B. Simon, On local
Borg-Marchenko uniqueness results. \textit{Commun.\ Math.\ Phys.}
{\bf 211} (2000) 273--287

\bibitem{Ishii} K.\ Ishii, Localization of eigenstates and
transport phenomena in one-dimensional disordered systems.
\textit{Progress Theor.\ Phys.\ Suppl.} {\bf 53} (1973), 77--118

\bibitem{Ishii/Matsuda} K.\ Ishii and H.\ Matsuda, Localization of
normal modes and energy transport in the disordered harmonic
chain. \textit{Progress Theor.\ Phys.\ Suppl.} {\bf 45} (1970),
56--86

\bibitem{Kato} T. Kato, \textit{Perturbation theory for linear
operators}. Springer-Verlag, Berlin--Heidelberg (1966)

\bibitem{Kirsch} W.\ Kirsch, On a class of random Schr\"{o}dinger operators.
\textit{Adv.\ in Appl.\ Math.} {\bf 6} (1985), 177--187

\bibitem{Kirsch/Kotani/Simon} W. Kirsch, S. Kotani and B. Simon: Absence
of absolutely continuous spectrum for some one dimensional random
but deterministic potentials. {\it Ann. Inst. Henri Poincar\'e}
{\bf 42}, 383 -- 406 (1985)

\bibitem{Kirsch/Stollmann/Stolz} W.\ Kirsch, P.\ Stollmann, and
G.\ Stolz, Anderson localization for random Schr\"odinger
operators with long range interactions. \textit{Commun.\ Math.\
Phys.} {\bf 195} (1998), 495--507

\bibitem{Klopp} F.\ Klopp, Localization for some continuous random
Schr\"odinger operators. \textit{Commun.\ Math.\ Phys.} {\bf 167}
(1995), 553--569

\bibitem{Kostrykin/Schrader} V.\ Kostrykin and R.\ Schrader,
Scattering theory approach to random Schr\"odinger operators in
one-dimension. \textit{Rev.\ Math.\ Phys.} {\bf 11} (1999),
187--242

\bibitem{Kostrykin/Schrader2} V.\ Kostrykin and R.\ Schrader,
Global bounds for the Lyapunov exponent and the integrated density
of states or random Schr\"odinger operators in one dimension.
Preprint 2000, mp-arc/00-226

\bibitem{Kotani} S.\ Kotani, One-dimensional Schr\"odinger
operators and Herglotz functions. In: \textit{Probabilistic Methods
in Mathematical Physics}, ed.\ K.\ Ito and N.\ Ikeda, 219--250,
Academic Press, Boston 1987

\bibitem{Li/Gr/Pa} I. Lifshits, S. Gredeskul, and L. Pastur,
Introduction to the theory of disordered systems (Translated from
the Russian by Eugene Yankovsky). \textit{A Wiley-Interscience
Publication}. John Wiley \& Sons, Inc., New York, 1988

\bibitem{Marchenko1} V. Marchenko, Certain problems in the theory
of second-order differential operators. \textit{Doklady Ahad. Nauk
SSSR} {\bf 72}, 457 --460 (1950) (Russian)

\bibitem{Marchenko2} V. Marchenko, Some questions in the theory of
one-dimensional linear differential operators of the second order.
I, \textit{Trudy Moskov. Mat. Obsc.} {\bf 1}, 327--420 (1952)
(Russian); English transl. in \textit{Amer. Math. Soc. Transl.
(2)} {\bf 101}, 1--104 (1973)

\bibitem{Figotin/Pastur} L.\ Pastur and A.\ Figotin,
\textit{Spectra of Random and Almost-Periodic Operators}. Springer
Verlag, Berlin-Heidelberg-New York (1992)

\bibitem{SVW} C.\ Shubin, R.\ Vakilian, and T.\ Wolff, Some harmonic analysis
questions suggested by Anderson-Bernoulli models. \textit{Geom.\ Funct.\ Anal.}
{\bf 8} (1998), 932--964

\bibitem{Simon1} B.\ Simon, A new approach to inverse spectral theory, I.
Fundamental formalism. \textit{Ann. of Math. (2)} {\bf 150} (1999)
1029--1057

\bibitem{st} B.\ Simon and M.\ Taylor, Harmonic analysis on SL$(2,\R)$ and smoothness
of the density of states in the one-dimensional Anderson model. \textit{Commun.\ Math.\
Phys.} {\bf 101} (1985), 1--19

\bibitem{Sims/Stolz} R.\ Sims and G.\ Stolz, Localization in one dimensional random
media: a scattering theoretic approach. Preprint 2000, mp-arc/00-64, to appear in
\textit{Commun.\ Math.\ Phys.}

\bibitem{PSW} P.\ Stollmann, Wegner estimates and localization for continuum Anderson
models with some singular distributions. Preprint 1998, to appear in \textit{Arch.\ Math.}

\bibitem{Stollmann} P.\ Stollmann, \textit{Caught by Disorder: Lectures on Bound States
in Random Media}. Book, in preparation.

\bibitem{Boundsol} G.\ Stolz, Bounded solutions and absolute continuity of Sturm-Liouville
operators. \textit{J.\ Math.\ Anal.\ Appl.} {\bf 169} (1992), 210--228

\bibitem{Stolz:Anderson} G.\ Stolz, Non-monotonic Random
Schr\"odinger Operators: The Anderson Model. \textit{J.\ Math.\
Anal.\ Appl.} {\bf 248} (2000), 173--183

\bibitem{Walter} W.~Walter, \textit{Ordinary Differential Equations}. Graduate Texts
in Mathematics, Vol.~{\bf 182}, Springer, New York (1998)

\end{thebibliography}
\end{document}